\documentclass[preprint,12pt]{elsarticle}

\usepackage{amsmath,graphicx,bm}
\usepackage[dvipsnames]{xcolor}
\usepackage{cite}
\usepackage[hidelinks]{hyperref}
\usepackage{multicol}
\usepackage{multirow}
\usepackage{todonotes}
\usepackage[tone,extra,safe]{tipa}
\newcommand{\ipa}[1]{\textipa{#1}}
\usepackage{booktabs}
\usepackage{amssymb}
\usepackage{subcaption}
\usepackage{comment}

\usepackage{amssymb}

\usepackage[acronym]{glossaries}
\makeglossaries
\newacronym{asr}{ASR}{Automatic Speech Recognition}
\newacronym{e2e}{E2E}{End-to-End}
\newacronym{ipa}{IPA}{International Phonetic Alphabet}
\newacronym{ma}{MA}{Main Aim}
\newacronym{per}{PER}{Phone Error Rate}
\newacronym{wer}{WER}{Word Error Rate}
\newacronym{rq1}{RQ1}{Research Question 1}
\newacronym{rq2}{RQ2}{Research Question 2}
\newacronym{rq3}{RQ3}{Research Question 3}
\newacronym{dnn}{DNN}{Deep Neural Network}
\newacronym{hmm}{HMM}{Hidden Markov Model}
\newacronym{auc}{AUC}{Area Under the Curve}
\newacronym{nmi}{NMI}{Normalized Mutual Information}
\newacronym{am}{AM}{Acoustic Model}
\newacronym{pl}{PL}{Pronunciation Lexicon}
\newacronym{lm}{LM}{Language Model}
\newacronym{ctc}{CTC}{Connectionist Temporal Classification}
\newacronym{umap}{UMAP}{Uniform Manifold Approximation and Projection}
\newacronym{jsd}{JSD}{Jensen-Shannon divergence}
\newacronym{g2p}{G2P}{Grapheme-to-Phone}
\newacronym{nlp}{NLP}{Natural Language Processing}
\newacronym{p}{P}{Phone}
\newacronym{pt}{PT}{Phone Token}
\newacronym{pter}{PTER}{Phone Token Error Rate}
\newacronym{tdnnf}{TDNNF}{Factorized Time-Delay Neural Network}
\newacronym{lfmmi}{LF-MMI}{Lattice-Free Maximum Mutual Information}
\newacronym{gmm}{GMM}{Gaussian Mixture Model}
\newacronym{mfcc}{MFCC}{Mel-Frequency Cepstral Coefficient}
\newacronym{tsne}{T-SNE}{T-distributed Stochastic Neighbor Embedding}
\newacronym{tp}{TP}{True Positive}
\newacronym{fp}{FP}{False Positive}
\newacronym{fn}{FN}{False Negative}
\newacronym{aos}{AoS}{Apraxia of Speech}
\newacronym{aoa}{AoA}{Age of Acquisition}
\newacronym{dof}{d.o.f.}{Degree of Freedom}

\journal{Computer Speech and Language}

\begin{document}

\begin{frontmatter}



\title{Discovering Phonetic Inventories with Crosslingual Automatic Speech Recognition}

\author[inst1,inst2]{Piotr~\.{Z}elasko}
\author[inst3]{Siyuan~Feng}
\author[inst1]{Laureano~Moro~Vel\'{a}zquez}
\author[inst4]{Ali~Abavisani}
\author[inst1]{Saurabhchand~Bhati}
\author[inst3]{Odette~Scharenborg}
\author[inst4]{Mark~Hasegawa-Johnson}
\author[inst1,inst2]{Najim~Dehak}

\affiliation[inst1]{organization={Center of Language and Speech Processing, The Johns Hopkins University},
            addressline={3400 North Charles Street}, 
            city={Baltimore},
            postcode={21218}, 
            state={MD},
            country={USA}}

\affiliation[inst2]{organization={Human Language Technology Center of Excellence, The Johns Hopkins University},
            addressline={810 Wyman Park Drive}, 
            city={Baltimore},
            postcode={21218}, 
            state={MD},
            country={USA}}

\affiliation[inst3]{organization={Multimedia Computing Group, Delft University of Technology},
            addressline={Van Mourik Broekmanweg 6}, 
            city={Delft},
            postcode={2628 XE},
            country={The Netherlands}}
            
\affiliation[inst4]{organization={Department of Electrical and Computer Engineering, University of Illinois},
            addressline={405 N Mathews}, 
            city={Urbana},
            postcode={61801}, 
            state={IL},
            country={USA}}

\begin{abstract}

The high cost  of data acquisition makes \acrfull{asr} model training problematic for most existing languages, including languages that do not even have a written script, or for which the phone inventories remain unknown.
Past works explored multilingual training, transfer learning, as well as zero-shot learning in order to build \acrshort{asr} systems for these low-resource languages. While it has been shown that the pooling of resources from multiple languages is helpful, we have not yet seen a successful application of an \acrshort{asr} model to a language unseen during training. A crucial step in the adaptation of \acrshort{asr} from seen to unseen languages is the creation of the phone inventory of the unseen language. 

The ultimate goal of our work is to build the phone inventory of a language unseen during training in an unsupervised way without any knowledge about the language. In this paper, we 1) investigate the influence of different factors (i.e., model architecture, phonotactic model, type of speech representation) on phone recognition in an unknown language; 2) provide an analysis of which phones transfer well across languages and which do not in order to understand the limitations of and areas for further improvement for automatic phone inventory creation; and 3) present different methods to build a phone inventory of an unseen language in an unsupervised way.
To that end, we conducted mono-, multi-, and crosslingual experiments on a set of 13 phonetically diverse languages and several in-depth analyses.
We found a number of universal phone tokens (\acrshort{ipa} symbols) that are well-recognized cross-linguistically. Through a detailed analysis of results, we conclude that unique sounds, similar sounds, and tone languages remain a major challenge for phonetic inventory discovery.


\end{abstract}



\begin{keyword}
phone inventory \sep \acrshort{asr} \sep speech recognition \sep multilingual \sep crosslingual \sep zero-shot \sep phone recognition \sep speech representation

\emph{Funding:} This work was funded by NSF grant number 1910319.
\end{keyword}

\end{frontmatter}


\section{Introduction}
\label{sec:intro}

\acrfull{asr} is one of the technologies deployed on a massive scale with the highest impact of the $21^{st}$ century. It has achieved enormous advancement during the past fifteen years~\citep{zhang2020Transformer}, enabling automatic transcription, translation, and the ubiquitous appearance of digital assistants, among other applications. Combined with spoken language understanding, \acrshort{asr} has the potential to automate numerous processes that require  spoken communication, such as requests for support and knowledge. Typically, the development of \acrshort{asr} systems requires hundreds to thousands of hours of speech recordings and their correspondent transcriptions. Unfortunately, only a small number of the world's languages is sufficiently resourced to build speech processing systems, while the amount of transcribed speech data for most of the 7,000 spoken languages in the world~\citep{austin2011cambridge} is very limited or nonexistent~\citep{speech2020scharenborg}.  This results in digital divides~\citep{Morrell2020} that can only be fixed with large investments in speech resources, if considering only traditional ways to develop \acrshort{asr} models for new languages.

One method that is often used to reduce the data requirements for under-resourced languages is to build word-level \acrshort{asr} out of phone models.  The word ``phone'' was coined in~\citep{lamel1989speech} as an abbreviation for ``phonetic symbol,'' defined in~\citep{leung1984procedure} as an element of a phonetic transcription corresponding to at most one phoneme, whose boundary times in the acoustic signal can be reliably identified using automatic forced alignment.   Such language-dependent \acrshort{asr} segment inventories may be expressed using the language-independent symbols of the \acrshort{ipa}~\citep{international1999handbook}, and their set union defines a language-independent phone inventory, which may be trained using multilingual data~\citep{schultz1998multilingual}; alternatively, language-dependent phone models may be trained using far less data than language-dependent word models, because the number of phones in a language is far fewer than the number of words~\citep{barnard2009asr}.  
In order to use phone-based acoustic models, however, it is necessary to discover the phone inventory of the unseen language.
Past research has addressed this problem and proposed knowledge transfer from high-resource to low-resource languages using \acrshort{asr} adaptation technologies. For instance, studies in~\citep{schultz1998multilingual,loof2009cross,Swietojansk2012unsupervised,Huang2013cross,li2020universal} propose  
multilingual (i.e., using training material from multiple languages including that of the target language) or crosslingual (i.e., using material from multiple languages excluding that of the target language) \acrshort{asr} systems, trained with several languages and targeting a low-resource target language. The reasoning behind the use of models that are trained with data from other languages is that sounds in different languages share phonetic representations due to similarities in articulatory and acoustic patterns. The phones of the target language are then assumed to be part of one or more of the training languages. However, this is not always the case, in which case those phones in the target language cannot be found. In some cases when a phone is not found in the training languages, it may be a subtle variant of some corresponding phones in the training languages, that could be used instead to form a phone system.

Most past research reports performance on the task of a phone or word transcription task in the unseen, target language. In this work, we go one step further and focus on the task of automatic, unsupervised phone inventory creation without any knowledge about the target language. To the best of our knowledge, we are the first to do so.

Below, we present our main aim and research questions. We will refer to them throughout the paper like this: \textbf{[\acrshort{ma}]}.

\textbf{Main Aim [\acrshort{ma}].} Our main research question is: Can the phone inventory of an unseen language be discovered in an unsupervised manner employing cross-lingual \acrshort{asr} systems trained with several languages? Previous research reports improvements in terms of \acrfull{per} or \acrfull{wer} when going from monolingual (low-resource languages) to multilingual training \citep{knill2013investigation,swietojanski2012unsupervised,Huang2013cross,li2020universal,dalmia2018sequence,swietojanski2012unsupervised,Swietojansk2012unsupervised}. Here, in addition to reporting \acrshort{per}, we investigate various factors that influence crosslingual \acrshort{asr} and automatic phone inventory creation for an unseen language, and investigate what the \acrshort{asr} models are learning in order to understand the limitations of and areas for further improvement for building \acrshort{asr} systems for low-resource languages, and particularly the creation of the phone inventory of an unseen language.

\begin{figure}
    \centering
    \includegraphics[width=\linewidth]{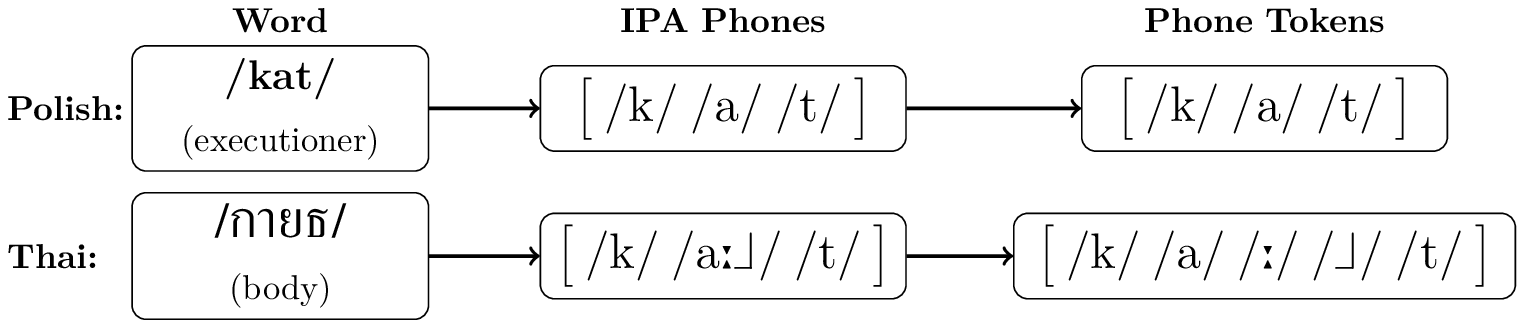}
    \caption{An overview of the basic phonetic units used in our study -- phones, which include modifiers such as tones as a part of the symbol, and phone tokens, where the modifiers are modelled separately. Both variants leverage \acrshort{ipa} symbols.}
    \label{fig:phonevsphonetoken}
\end{figure}

\textbf{Research Question 1 [\acrshort{rq1}].}
Firstly, we investigate what types of speech representations to use to train our models. Phonemes are language-dependent sound categories whose substitution may change the meaning of a word~\citep{swadesh1934phonemic}.  Phones are the acoustic realizations of  phonemes whose boundary times in the acoustic signal can be reliably identified using automatic forced alignment.
The number of non-phonemic contrasts in  the \acrshort{asr} segment inventory is small in some languages, e.g., a  typical American English \acrshort{asr} distinguishes a few allophones of \ipa{/t/}, such as \ipa{[R]} and~\ipa{[t]}~\citep{Lee1989}.  In other \acrshort{asr} segment inventories, the number of suprasegmental contrasts may be quite large, e.g., a typical \acrshort{asr} segment inventory for Mandarin Chinese distinguishes among all possible tonally-marked syllable finals~\citep{lin1993new}:
  $[$a\tone{55}$]$,
  $[$\textipa{a}\tone{33}\tone{55}$]$,
  $[$\textipa{a}\tone{22}\tone{11}\tone{44}$]$,
  $[$\textipa{a}\tone{55}\tone{11}$]$,
  and the neutral-tone vowel
  $[$\textipa{a]}
  are all distinct.
  Such language-dependent \acrshort{asr} segment inventories may be expressed using the language-independent symbols of the \acrshort{ipa}~\citep{international1999handbook}, and their set union defines a universal (language-independent) phone inventory.  We define the goal of our research to be the discovery of the language-dependent \acrshort{asr} segment inventory as a subset of a larger, multilingual phone inventory.  Defining our goal in this way, however, creates some ambiguity: if Mandarin \acrshort{asr} uses
  $[$\textipa{a}\tone{22}\tone{11}\tone{44}$]$
  as an element in its standard \acrshort{asr} segment inventory, but our phonetic inventory discovery system discovers only the phoneme \textipa{/a/},
  should that be considered a correct, incorrect, or partially correct result?  The ambiguity is solved, in this paper, by defining two levels of language-independent annotation.  Our {\em phone} inventory is the strict set union of the language-dependent \acrshort{asr} segment inventories, expressed in \acrshort{ipa} notation.  Our {\em phone token} inventory is the set of all unicode characters used to express the phone inventory.
  Thus for example, the phone inventory includes the distinct phones
  $\{\ldots$,
  \textipa{a},
  \textipa{a}\tone{55},
  \textipa{a}\tone{33}\tone{55},
  \textipa{a}\tone{22}\tone{11}\tone{44},
  \textipa{a}\tone{55}\tone{11},
  $\ldots\}$,
  while the phone token inventory includes the distinct phone tokens
  $\left\{\ldots\right.$,
  \textipa{a},
  \tone{11},
  \tone{22},
  \tone{33},
  \tone{44},
  \tone{55},
  $\left.\ldots\right\}$.
We distinguish two types of  phone tokens: symbols that describe acoustic segments (base phones such as~\ipa{[a]}), and symbols that modify the base phone in order to produce a different phone (diacritics, including  secondary articulations, lengthening, lexical stress, and lexical tone, such as [\tone{55}]).
Diacritics such as the high tone symbol [\tone{5}] have no sound unless coupled with a base  phone such as  [\textipa{a}], but perhaps cross-lingual \acrshort{asr} can be learned more effectively by a  system that decomposes [\textipa{a}\tone{5}] into its  component phone  tokens, i.e., a system where [\textipa{a}]
and [\tone{5}] are modelled separately (Figure~\ref{fig:phonevsphonetoken}).
Therefore, we aim to answer the question: Can phone-based or phone-token-based acoustic models, trained using data from several languages coded with the same \acrfull{ipa} symbols, be used to detect the presence of a phone coded with the same \acrshort{ipa} symbol(s) in the \acrshort{asr} segment
inventory of a different language?

\textbf{[\acrshort{rq2}].} Secondly, we know that the speech models learn phonetic representations~\citep{nagamine2015exploring}, however, it is unclear: Which phones' representations are the most, and the least, amenable to crosslingual transfer? We also look for the factors that may help in explaining the findings. Using several in-depth analyses we answer these questions.

\textbf{[\acrshort{rq3}].} Thirdly, it is not possible to have a complete picture on how multilingual representations can be used in zero-shot scenarios without studying to what degree the mismatch of phonotactics\footnote{Phonotactics define restrictions in a language on the permissible combinations of phonemes.} between training and target languages affect the overall phone recognition results of the target language.
Consequently, our third research question is: How do phonotactic constraints influence phone recognition in multilingual and crosslingual scenarios? Is this influence also visible in phonetic inventory discovery performance?

\textbf{Proposed solution.} To answer these questions, we trained two architectures with different properties, i.e., an \acrfull{e2e} phone-level \acrshort{asr} and a hybrid \acrshort{dnn}-\acrshort{hmm} \acrshort{asr}, using training data from 13 languages from 8 language families and with vastly different phonetic properties. This experimental protocol is based on the preliminary exploration we have performed before -- in~\citep{Zelasko2020That}, we looked into \acrshort{e2e} \acrshort{asr} phonetic representation transfer across languages, and in~\citep{feng2020phonotactics} we investigated the effect of phonotactics on zero-shot \acrshort{asr} transfer.

Hybrid \acrshort{asr} systems explicitly factorize the acoustic and language models into independent modules, which allows us to quantify the influence of the language phonotactics on phone recognition performance in multilingual and zero-shot scenarios (\textbf{[\acrshort{rq3}]}).
To measure the effect of phonotactic information, phonotactic \acrshort{lm}s of different capacities were trained and compared alongside the hybrid \acrshort{asr}.
In our \acrshort{asr} experiments, we used both \acrshort{ipa} phones and phone tokens as our speech representations (\textbf{[\acrshort{rq1}]}). In our experiments, we use three scenarios: 
monolingual, as a baseline; multilingual, to investigate the usefulness of pooling all languages together; and crosslingual, or zero-shot \acrshort{asr}.
In each scenario, we evaluate the degree to which the confusion matrix  rows corresponding to different phones, phone tokens, or classes of phones are helped or harmed by multilingual or cross-lingual  modeling  (\textbf{[\acrshort{rq2}]}).  
Finally, in order to address the \textbf{Main Aim [\acrshort{ma}]} of the paper,
we employed the best \acrshort{e2e} and hybrid architectures in a crosslingual scenario, and propose and compare several methods for automatic phone inventory discovery. 

\section{Related work}
\label{sec:related_work}

\textbf{Phonetic inventory discovery.}  The language-dependent categorical perception of phonemes was strongly argued in a series of thought experiments by linguists including Trubetzkoy~\citep{trubetzkoy1929zur} and Sapir~\citep{sapir1925sound}.  While trying to define the phoneme inventory of the native American language Chitimacha, Swadesh provided one of the clearest published definitions of a phoneme: ``The phoneme is the smallest potential unit of difference between similar words recognizable as different to the native.  Given a  correct native word, the replacement of one or more phonemes by other phonemes (capable of occurring in the same position) results in a native word other than that intended, or a native-like nonsense word''~\citep{swadesh1934phonemic}.  During the following twenty years, the Swadesh principle was used to propose phoneme inventories for hundreds of languages, including English~\citep{trager1941syllabic}, Russian~\citep{trager1934phonemes}, Amahuaca~\citep{osborn1948amahuaca}, Cashibo~\citep{shell1950cashibo}, Huasteco~\citep{larsen1949huasteco}, Totonaco~\citep{aschmann1946totonaco}, and Zoque~\citep{wonderly1951zoque}.  The PHOIBLE database~\citep{Moran2019} lists 3020 phoneme inventories that have been published as descriptions of 2186 languages.

There is a great deal of research on the automatic discovery of phone inventories for downstream technological tasks.  Four types of system evaluation are prevalent.  First, systems participating in the zero resource speech recognition challenges~\citep{versteegh2015zero}, including~\citep{Tjandra2019,pellegrini2017technical,heck2017feature,shibata2017composite,ansari2017deep,ansari2017unsupervised,Hermann2018,hermann2018multilingual_journal,wu2018optimizing,renshaw2015comparison,Heck+2016,heck2016unsupervised,Heck2016iterative,tsuchiya2018speaker,Bhati2017} are evaluated using an ABX task similar to the Swadesh minimal-pairs test~\citep{swadesh1934phonemic}, but are not asked to propose a phone inventory whose speech sound clusters match to any human-annotated ground truth.  Second, unsupervised phone discovery systems may be evaluated as components of zero-resource keyword spotting~\citep{menon2019feature,levin2015segmental,9004019,siu2010improved,WangRagniGalesEtAl2015,SiuGishChanEtAl2014,ni2016rapid}.  Third, unsupervised phone discovery systems may be evaluated for the purpose of developing automatic speech synthesis in a zero-resource language~\citep{muthukumar2014automatic,baljekar2015using,muller2017towards,Dunbar2019,yu2016learning}.  Fourth, unsupervised phone discovery can be used as a component in unsupervised lexicon discovery~\citep{lee2015unsupervised,bacchiani1999joint,kamper2016unsupervised} or unsupervised speech-to-text~\citep{baevski2021unsupervised}.  Unsupervised  speech-to-text may be evaluated on the  basis of word error rate or phone error rate, but has not, to our knowledge, been evaluated in the task of phone inventory discovery.

Comparison of an automatically discovered phone inventory to a ground-truth, human annotated phone inventory has been performed in certain limited ways. Kempton et al.~\citep{Kempton2014} force-aligned human-annotated fine phonetic transcriptions to audio in an under-resourced language (Kua-nsi), and then evaluated the degree to which the force-aligned transcriptions could be clustered to compute the true phoneme inventory of the language.  Results indicated the need for further improvements for any particular phonetic class, e.g., active articulator was correctly classified with only 0.745 \acrfull{auc}.  Following  up  on their study, several researchers have attempted unsupervised clustering of acoustic spectra using Bayesian~\citep{lee2012a}, variational~\citep{ondel2016variational} or hidden Markov models~\citep{Ondel2019Bayesian,Ebbers2017}, in order to directly  estimate the phone inventory of an unknown language.  The cardinality of the resulting pseudo-phone  set is usually ten times  larger than the cardinality of the true phone inventory, but by associating each pseudo-phone to its most highly overlapped ground-truth phone, it is possible to evaluate the pseudo-phone inventory using metrics such as F1 and normalized mutual information (\acrshort{nmi},~\citep{estevez2009normalized}).  Finally, several recent studies have attempted a combination of Kempton's approach with the unsupervised clustering approach: cross-lingual \acrshort{asr} is used to annotate the articulatory features of an unknown language, which are then clustered to form unsupervised phone-like units~\citep{muthukumar2014automatic,baljekar2015using}.  To our knowledge, only two of these papers~\citep{yusuf_ondel,feng2021} directly evaluated phone inventory \acrshort{nmi} or F1; using oracle cluster combination strategies that are standard in the field of unsupervised phone discovery,~\citep{feng2021} achieved F1=64.14\% for cross-lingual automatic phone inventory estimation.

\textbf{Crosslingual \acrshort{asr}.} We call an \acrshort{asr} system that may recognize more than one language as \emph{multilingual}; when the target language is not seen during training, then the system becomes \emph{crosslingual}.  Multilingual \acrshort{asr} models have been a point of interest even before \acrfull{dnn} became the state-of-the-art acoustic modelling technique~\citep{schultz1998multilingual}. Pooling multiple languages into the training set -- with or without architectural changes in the neural network-based acoustic models -- is known to be effective for improving \acrshort{asr}~\citep{schultz1998multilingual,SCHULTZ200131,NIESLER2007453,Swietojansk2012unsupervised,knill2013investigation,Huang2013cross,ghoshal2013multilingual,dalmia2018sequence,toshniwal2018Multilingual,kannan2019large,li2020universal}. Note that both monolingual and multilingual systems can be applied crosslingually, although multilingual systems can be expected to provide a more universal phonetic representation~\citep{Zelasko2020That}. Two major challenges specific to crosslingual \acrshort{asr} are \emph{mismatched outputs}, as both graphemic and phonemic symbols are usually different across languages; and \emph{mismatched inputs}, as the new languages might consist of sounds unseen in the training languages. Past works address this challenge with attempts to adapt the training vocabularies to new languages~\citep{byrne2000towards,bruguier2017pronunciation,patel2018cross}, convert the lexicons of other languages to the target language~\citep{yu2009cross,9004019}, or extend the vocabulary and fine-tune on the target language's data~\citep{tong2017multilingual,tong2018cross,inaguma2019transfer}. Another recent direction is pre-training the models to learn word-level acoustic representations~\citep{hu2020multilingual,9383545,9383594}.  Apart from our proposed \acrshort{asr} systems, any \acrshort{asr} system attempting to recognize language-independent (i.e., phonetic) units, such as the multilingual allophone approach proposed in~\citep{li2020universal}, would be suitable for performing phonetic inventory discovery.


\textbf{Analysis of learned representations.} Most of the mentioned studies do not perform a detailed analysis of which phones, manner and place of articulation classes, or language families can benefit more from multilingual and crosslingual training. This knowledge could be essential to build better \acrshort{asr} systems for languages with low or no resources. 
Finally, to our knowledge, no study performs \acrshort{ipa} phone inventory discovery of unknown languages by exclusively employing cross-lingual \acrshort{asr} systems, without additional acoustic clustering based on audio in the target language.  By studying this problem, we are able to make statements about the degree to which phonemes and phones may be approximated by universal, language-independent categories.

\section{Methods}
\label{sec:methods}

In this section we present the details of our approach to phonetic inventory discovery. We start with a brief revision of the architectures of two \acrshort{asr} system types used in our study: the hybrid \acrshort{asr} in Section~\ref{sec:methods:hybrid}, and the end-to-end \acrshort{asr} in Section~\ref{sec:methods:e2e}. We use two \acrshort{asr} toolkits: Kaldi\footnote{\url{https://github.com/kaldi-asr/kaldi}} and ESPnet\footnote{\url{https://github.com/espnet/espnet}}. In Section~\ref{sec:methods:confusion} we propose a novel method of discovering which phones are similar to each other across languages, based on clustering and projection of the \acrshort{asr} phonetic confusion matrix. Finally, we describe our method of discovering phonetic inventories from \acrshort{asr} transcripts in Section~\ref{sec:methods:disco}.



\subsection{Hybrid DNN-HMM ASR}
\label{sec:methods:hybrid}

A hybrid \acrshort{asr} system allow us to inspect the impact of phonotactics on the crosslingual system performance and phonetic inventory discovery \textbf{[\acrshort{rq3}]}. 
The hybrid system consists of three independent modules. 
The first is the \acrfull{am}, which predicts the posterior likelihood $p(\bm{X} | \bm{S})$ of phones or phone \acrshort{hmm} states $\bm{S} = \{s_0, s_1, ... s_N\}$ given the acoustic feature frames $\bm{X} = \{x_0, x_1, ..., x_N\}$.
Modern acoustic models are usually implemented with deep neural networks.
The second module is the \acrfull{pl} that maps sequences of phones into words.
The third module is the \acrfull{lm}, which predicts the conditional likelihood  $p(w_{i+1} | w_0, w_1, ... w_i)$ of the next word $w_{i+1}$ given the previous words.
All these components are incorporated together in a weighted finite-state transducer framework~\citep{mohri2002weighted}, and the most likely word sequence is retrieved using graph search methods, such as beam search. 
For more details, we refer the reader to~\citep{mohri2002weighted} and the documentation of the Kaldi \acrshort{asr} toolkit\footnote{https://kaldi-asr.org/doc/}.

The above description concerns word-level hybrid \acrshort{asr} systems. 
The modifications we apply to this framework are the following:
\begin{enumerate}
    \item The \acrshort{pl} is only applied during training to convert the word-level transcripts into phone sequences. During decoding, we are only interested in knowing phonetic transcripts.
    \item For decoding, we replace the word-level \acrshort{lm} with a phone-level \acrshort{lm} when converting the \acrshort{am} posteriors into a sequence of phones. We further refer to the phone-level \acrshort{lm} as the phonotactic \acrshort{lm}, or the phonotactic model.
\end{enumerate}

Note that we do not experiment with phone-token hybrid \acrshort{asr} systems \textbf{[\acrshort{rq1}]}. 
Hybrid \acrshort{asr} depends on high quality frame-level alignments during its training. 
We believe that the task of aligning symbols such as tone modifiers is ill-defined, as they are a property of the base phone and not a separate sound that follows the base phone.
However, the same limitation is not shared by another class of \acrshort{asr} systems we are about to discuss.

\subsection{End-to-End ASR}
\label{sec:methods:e2e}

In the recent years, another class of \acrshort{asr} systems, popularly called \acrfull{e2e}, has grown as an alternative to the hybrid systems. These systems will help us investigate which output label type, phones or phone tokens, could be more beneficial for crosslingual \acrshort{asr} \textbf{[\acrshort{rq1}]}. 

Whereas the exact definition of \acrshort{e2e} is not clear, typically this term describes models that perform graphemic transcription directly, i.e., they require neither an additional pronunciation lexicon, nor even an external language model to transcribe\footnote{Note that a fusion with an external language model can still improve \acrshort{e2e} system's performance.}. 
These systems are trained without using a forced or ground-truth alignment -- often using a \acrfull{ctc} loss, or an attention encoder-decoder architecture. 
The \acrshort{ctc} objective is frame-synchronous (i.e., each frame is assigned a label) and maximizes the likelihood of observing the correct transcript given the acoustic frames over all possible alignments corresponding to the ground truth text~\citep{graves2006connectionist}.

The encoder-decoder attention model consists of two modules: an encoder, and a decoder with attention. 
The encoder first encodes the input frame sequence (possibly sub-sampling it first), and produces a sequence of intermediate representations that are used as decoder inputs. 
The decoder is autoregressive - it models the conditional probability of the next output token, conditioned on the previous output token and the intermediate representations provided by the encoder. 
The first decoding step is initialized with a special $[BOS]$ symbol (\emph{beginning-of-sentence}), and it stops once the decoder recognizes a special $[EOS]$ symbol (\emph{end-of-sentence}). 
For each prediction step, the encoder's activations are aggregated via a weighted sum, with the weights determined by the source-target attention mechanism~\citep{chorowski2015attention}.
The decoder's output sequence is typically much shorter than the vector of acoustic frames, making encoder-decoder a frame-asynchronous model.
While the encoder-decoder attention model has the advantage of modelling the acoustics and language jointly, it is susceptible to misalignment due to noise or convergence issues. Hence, \citep{watanabe2017language} and \citep{hori2017joint} proposed to combine the \acrshort{ctc} and encoder-decoder attention model to take advantage of both systems' capabilities. Going forward, in this work we will refer to this joint \acrshort{ctc} and attention system as \acrshort{e2e}. Note that whereas usually \acrshort{e2e} means a system that performs graphemic transcription, we are specifically using the same approach to generate an \acrshort{ipa} phonetic  transcription, using units  of  either  phone  tokens (each decoder output generates one \acrshort{ipa} character) or phones (each decoder output generates one or more \acrshort{ipa} characters).

Note that the \acrshort{e2e} system does not allow us to inspect the impact of the phonotactics \textbf{[\acrshort{rq3}]} as they are learned jointly with the acoustic representations (even when we do not use an external \acrshort{lm}).

\subsection{Visualizing phonetic confusions}
\label{sec:methods:confusion}

\begin{figure*}[hbpt!]
    \centering
    \includegraphics[width=\textwidth]{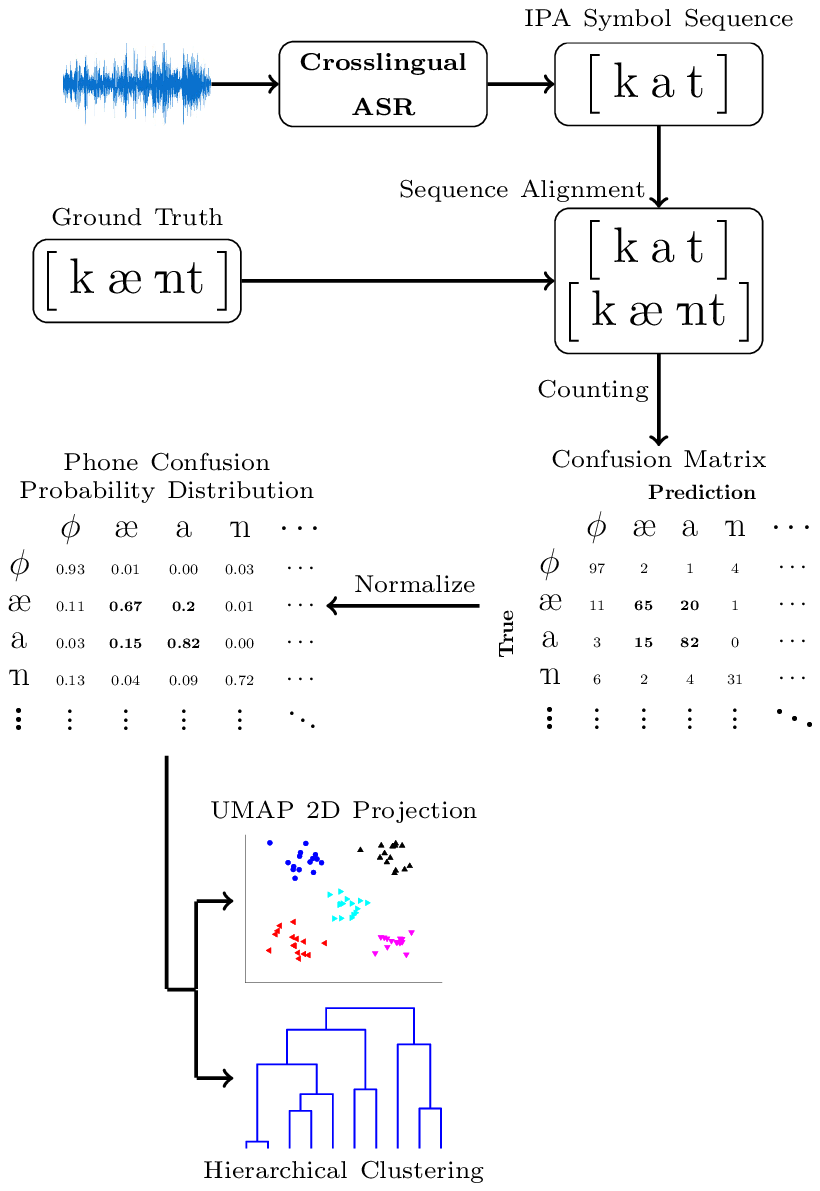}
    \caption{High-level workflow illustrating how the phone confusion matrix is obtained and mapped into phone similarity visualisations.}
    \label{fig:confmatrixvis}
\end{figure*}

We aim to learn which phones' representations are the most amenable to crosslingual transfer and which phones are hard to recognize crosslingually \textbf{[\acrshort{rq2}]}. To that end, we analyze the patterns in the zero-shot systems' phone confusions. A phone (token) confusion can be interpreted as: a given target-language phone (token) resembled a different source-language phone (token) more than it resembled the same source-language (``correct'') phone (token).  The cross-lingual confusion matrix therefore represents the degree to which each phone or phone token fails to transfer well across languages.

Our approach is visualized in Figure~\ref{fig:confmatrixvis}. We compute the edit distance alignments between the reference and hypothesized phone sequences and use them to estimate the phone confusion matrix.
Unfortunately, the confusion matrix is quite large due to the size of the \acrshort{ipa} symbol vocabulary, making a direct inspection of it hardly feasible.
Instead, inspired by some of the methods of confusion matrix re-ordering~\citep{behrisch2016matrix}, we transform it to the following two representations: a hierarchical clustering dendrogram, and a 2-D \acrfull{umap}~\citep{mcinnes2018umap}. The dendrogram allows us to easily view specific phone token clusters, whereas the projection hints at the global structure of the confusion space and the relationships between different symbols and clusters.

The clustering and projection algorithms both depend on the estimation of a distance matrix with a carefully selected distance metric.
In order to obtain meaningful visualisation, we propose to view the confusion matrix as a set of probability distributions -- for each phone token (row), what is the probability of confusing it with another phone token (column).
The intuition behind this approach is that when two symbols tend to be confused with the same set of classes, they must be similar to each other.
We find it helpful to first zero out the confusion pairs that occurred less than 50 times, and then remove the phone symbols that have empty (all zero) rows or columns.
Using this threshold, we reject about 5.3\% of all confusions (452202 out of 8536116 symbols confused in the transcripts), which account for approx. 37\% of all confusion types (e.g., we understand confusing [a] with [e] and [a] with [o] as a different confusion type; we discard 21168 out of 57120 confusion types). Without any thresholding, the patterns supported by most of the data are obfuscated by a long tail of noisy patterns after the occurrence counts are normalized.
Then, we preprocess the confusion matrix with row-normalization, so that each row is converted from occurrence counts to a categorical probability distribution of confusing a ground truth phone $P$ with another phone $P'$. 
We are going to interpret these distributions as feature vectors, describing the similarity between symbol pairs.
Representing each phone as a probability distribution provides us a sound motivation to select the \acrfull{jsd} as the distance metric.

\subsection{Phonetic inventory discovery}
\label{sec:methods:disco}

\begin{figure*}[hbpt!]
    \centering
    \includegraphics[width=\textwidth]{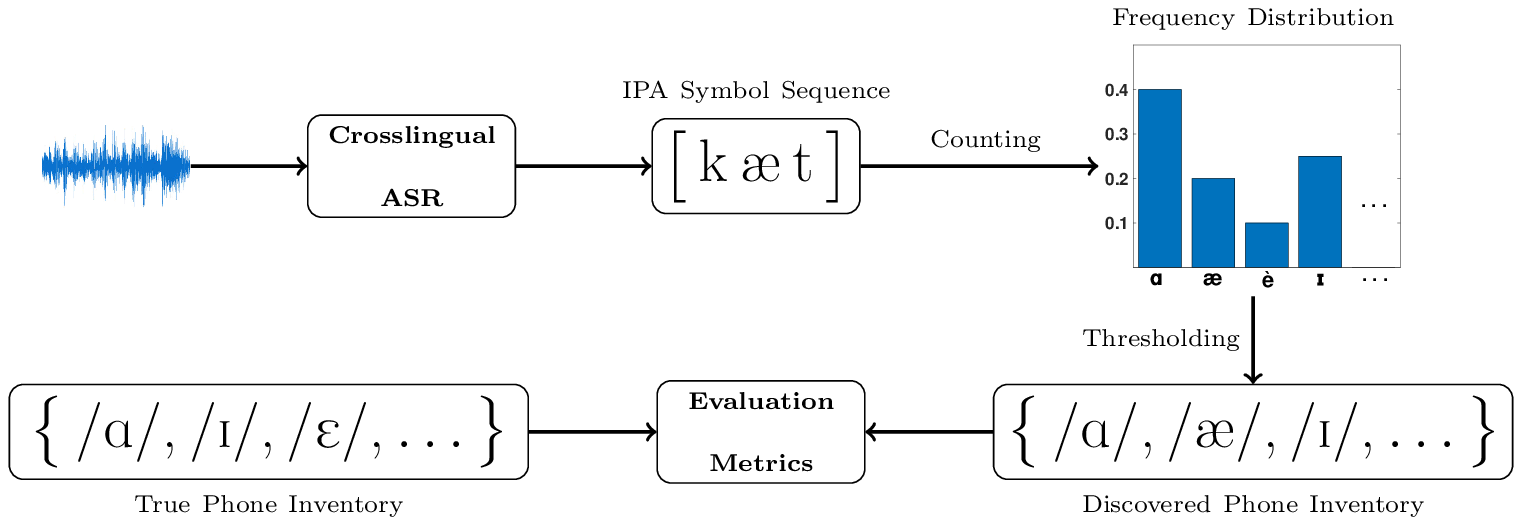}
    \caption{High-level workflow illustrating how a crosslingual (zero-shot) \acrshort{asr} is applied to phone inventory discovery of an unknown language.}
    \label{fig:discophonediag}
\end{figure*}

The ultimate goal of our works is to automatically discover the phone inventories of an unknown language \textbf{[\acrshort{ma}]}. Note that in this paper, we usually mention \textbf{phonetic inventories}. We use that term, in a slight abuse of terminology, to refer to any of the two different types of inventories -- a \textbf{phone inventory} and a \textbf{phone token inventory}.
The \textbf{phone inventory} of a language is the set of phones generated by a language-dependent \acrfull{g2p}.  The \textbf{phone token inventory} is the set of \acrshort{ipa} characters that make up those phones.

\textbf{Frequency-based inventory discovery.} To probe the zero-shot system's usefulness in phonetic inventory discovery, we propose a straightforward approach: we assume that all symbols that occur ``frequently enough'' in the zero-shot system's output for a given target language together constitute the target language's phonetic inventory. But what is ``frequently enough''? Here, we simply count the number of occurrences of each symbol in the recognized transcripts of an unseen language and set a threshold, where any symbol with a frequency above the threshold is considered to be part of the language's phonetic inventory.
To ensure that the approach is not strongly dependent on the amount of available test data, we convert the absolute occurrence counts into relative frequencies, dividing each count by the total number of recognized symbols for that language.

Phone inventory discovery cannot be performed using a phone-token \acrshort{asr}: there are many different ways in which the phone tokens generated by an \acrshort{asr} could be recombined into phones, and the different combination methods have different phone discovery accuracies. 
Both types of inventory discovery (phone inventory and  phone token inventory) can be performed using phone-based \acrshort{asr}: the only difference between them is that, for phone token discovery, the phones in the \acrshort{asr} output are split into phone tokens prior to application of the process in Figure~\ref{fig:discophonediag}.

\section{Experimental setup}
\label{sec:exp}

This section presents the details of our experiments. 
We start by introducing the corpora and languages used in our study in Section~\ref{sec:exp:corpora}.
Section~\ref{sec:exp:asr} describes the setup of our 
monolingual, multilingual and crosslingual \acrshort{asr} experiments.
Finally, we describe the phonetic inventory experiments in Section~\ref{sec:exp:disco}.

\subsection{Corpora}
\label{sec:exp:corpora}

 \begin{table}[!t]
\renewcommand\arraystretch{0.7}
\centering
\caption{Speech data used in the experiments. Train and Eval columns are presented in the number of hours. Vow and Con indicate the number of distinct phones used to describe vowels and consonants, respectively. Uniq denotes the number of \acrshort{ipa} phones only existing in that language. The symbol $^{\ddagger}$ indicates tone languages.}
\resizebox{ 0.99 \linewidth}{!}{%
\begin{tabular}{c|l|ccccc}      
\toprule
Corpus & Language & Train & Eval & Vow & Con & Uniq \\
\midrule
\multirow{5}{*}{GlobalPhone} &
Czech & 24.0 & 3.8 & 10 & 25 &5\\
&French & 22.8 &  2.0 &19&26&9\\
&Spanish & 11.5 &  1.2 &6&24&4\\
&Mandarin$^{\ddagger}$ & 14.9 & 1.6&60&33&41\\
&Thai$^{\ddagger}$ & 22.9 & 0.4&108&35&17\\
\midrule
\multirow{8}{*}{Babel} &
Cantonese$^{\ddagger}$ &126.6 & 17.7 &110 &25&88\\
&Bengali & 54.5 & 9.8&19&33&13\\
&Vietnamese$^{\ddagger}$ & 78.2 & 10.9&204 &25&130\\
&Lao$^{\ddagger}$ & 58.7 & 10.5&106&61&83\\
&Zulu$^{\ddagger}$ & 54.4 & 10.4&15&54&19\\
&Amharic &  38.8 & 11.6 &8& 52&9\\
&Javanese & 40.6 & 11.3 &11&23&1\\
&Georgian &  45.3 & 12.3 &5&30&2\\
\bottomrule
\end{tabular}%
}
\label{tab:database}
\end{table}

We selected $13$ languages for this study which are shown in Table \ref{tab:database}, identical to those used in our previous work~\citep{Zelasko2020That,feng2020phonotactics}. These languages were chosen for the diversity of their phoneme inventories. 
For instance, Bengali was chosen as a representative language with voiced aspirated stops, Javanese for its slack-voiced (breathy-voiced) stops, Zulu has clicks, and Mandarin, Cantonese, Vietnamese, Lao, and Thai are tone languages with very different tone inventories (Zulu also has lexical tones, but our grapheme-to-phoneme transducer for Zulu does not label the tones of each syllable). Table~\ref{tab:database} reports the number of distinct phones in the \acrshort{asr} segment
inventory for that language.  Each phone  consists of a base phone  (a speech sound associated with a temporal segment), possibly modified by diacritics marking secondary articulation, lengthening, lexical stress, or lexical tone, e.g., the high-, rising-, dipping- and falling-tone /i/ in Mandarin are described by four different phones [i\tone{55}], [i\tone{33}\tone{55}], [i\tone{22}\tone{11}\tone{44}], and [i\tone{55}\tone{11}], respectively.  Phones may be affricates if affricates are phonemic in the source language, e.g., the five Amharic phones~\textipa{/t/},\textipa{/t'/},\textipa{/S/},\textipa{/tS/},\textipa{/tS'/} are coded using only three \acrshort{ipa} symbols:~\textipa{[t]},~\textipa{[S]}, and~\textipa{[']}. Note that the high number of vowels in tone languages is due to different combinations of tone modifiers (e.g., Vietnamese and Cantonese each have six tones per vowel).
We obtained the \acrshort{ipa} phonetic transcripts by leveraging the LanguageNet \acrfull{g2p} models~\citep{hasegawa2020grapheme}.

Speech data for these languages were obtained from the GlobalPhone~\citep{schultz2002globalphone} and IARPA Babel corpora. GlobalPhone has a small number of recording hours for each language - about 10 to 25 hours - and represents a simpler scenario with limited noise and reverberation. On the other hand, Babel languages have more training data, ranging from 40 to 126 hours, but these databases are significantly more challenging due to the recordings having been made in a larger variety of acoustic environments. We used standard train, development, and evaluation splits for Babel. For GlobalPhone languages, whenever the documentation provides standard split information, we use these -- otherwise, we chose the first 20 speakers' recordings for development and evaluation sets (10 speakers each), and train on the rest. Table 1 gives an overview of the number of hours of training and test (Eval) material for each language.

\subsection{ASR experiments}
\label{sec:exp:asr}

\begin{figure}
    \hspace*{2cm}
    \includegraphics[width=.6\textwidth]{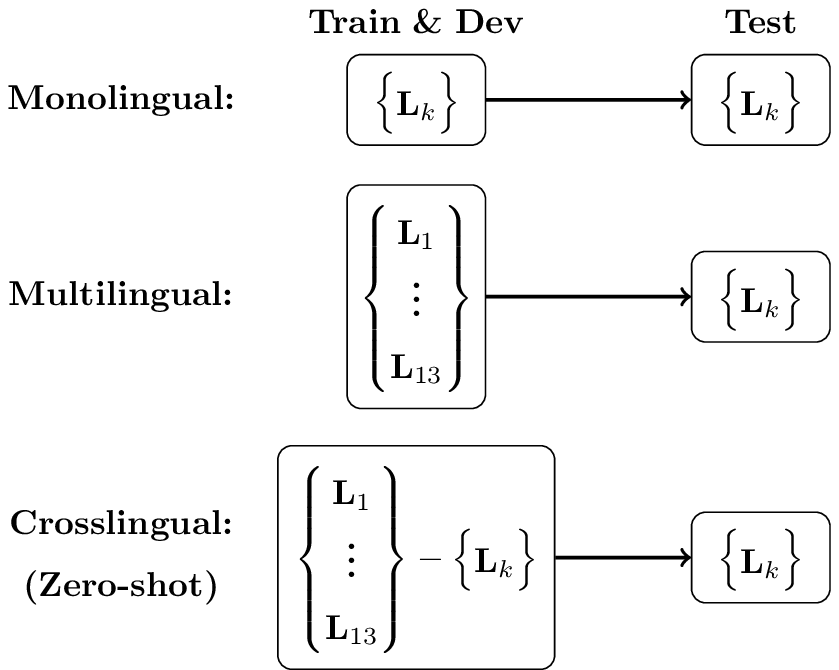}
    \caption{Three \acrshort{asr} experimental scenarios in our study: monolingual, multilingual, and crosslingual. $L_k$ indicates $k$-th language in our experimental dataset.}
    \label{fig:monomulticross}
\end{figure}

The ultimate aim of our work is to discover phonetic inventories \textbf{[\acrshort{ma}]}. 
In order to be able to asses the strengths and weaknesses of the crosslingual \acrshort{asr} approach, we also need baseline systems that have learned a robust phonetic representation of the target language. They will serve as oracles for evaluating phonetic inventory discovery. Hence, our \acrshort{asr} experiments are divided into three scenarios (mimicking our previous setups~\citep{Zelasko2020That,feng2020phonotactics}; see also Figure~\ref{fig:monomulticross}):
\begin{enumerate}
    \item monolingual (mono) -- 13 different \acrshort{asr} systems are trained and tested each on the same language;
    \item multilingual (multi) -- a single \acrshort{asr} is trained and tested on all languages; and
    \item crosslingual (cross) -- 13 different \acrshort{asr} systems are each trained on 12 languages and tested on a held-out language. We also refer to this system as a \textbf{zero-shot} \acrshort{asr}, as the target language is never seen in training.
\end{enumerate}

\textbf{\acrshort{e2e} \acrshort{asr} experiments.} 
We used an \acrshort{e2e} \acrshort{asr} system whose inputs were combined filterbank and pitch features~\citep{ghahremani2014pitch} with a frame length of 25ms and a frame shift of 10ms. The \acrshort{asr} was trained with joint \acrshort{ctc} and attention objectives~\citep{kim2017joint}, employing the ESPnet toolkit~\citep{watanabe2018espnet}. The setup is based on the system described in~\citep{karita2019comparative} that uses a transformer architecture. Transformer-based models have been shown to produce state-of-the-art results on several \acrfull{nlp} tasks~\citep{vaswani2017attention,devlin2018bert}, and has recently been adopted in speech processing~\citep{karita2019comparative}.
We use $12$ transformer layers for the encoder and $6$ for the decoder. 
Each transformer  layer consists of $4$ attention heads with a dimension of $256$ and $2048$-dimensional feed-forward layer.
Prior to the transformer layers, the encoder uses $2$ 2D convolutional layers with a stride of $2$ each to sub-sample the signal with a factor of $4$.
We use label smoothing with a factor of $0.1$.
To accommodate the transformer's $O(n^2)$ memory requirements, we discard all training and test utterances which exceed 30 seconds of speech (about 1.7\% of utterances). 

Two types of \acrshort{e2e} experiments were carried out, one using \acrshort{ipa} phones as speech representation (\acrshort{e2e} \acrshort{p}) and one which used phone tokens as speech representation (\acrshort{e2e} \acrshort{pt}) \textbf{[\acrshort{rq1}]}.
The \acrshort{e2e} \acrshort{p} output alphabet is the union of \acrshort{ipa} phones in all training languages, hence it cannot recognize novel target-language phones in a  crosslingual setting.  The \acrshort{e2e} \acrshort{pt}  output alphabet is the union of all unique unicode characters representing \acrshort{ipa} base phones or diacritics in any training language.  It is not always clear how we should segment the output  of the \acrshort{e2e} \acrshort{pt}  system in order to measure phone error rate.
We take this phenomenon into account during the evaluation, by measuring \textbf{\acrfull{pter}} rather than \acrfull{per}.  \acrshort{pter} treats every modifier symbol as a separate token. 
That means that if the reference transcript has a vowel with a tone symbol, and the \acrshort{asr} recognized the vowel correctly but omitted the tone, we would count this as one correct token and one deletion instead of a single substitution~\footnote{
Note that we introduced PTER in our best effort to make P and PT systems comparable. The only alternative we see is imposing a decoding constraint on the E2E PT system that forces it to choose paths that only contain allowed phone token sequences that correspond to phones in an inventory known up-front. That is however equivalent to introducing a language model and FST decoding. We decided not to pursue LM integration with E2E models for the sake of avoiding additional complexity.}.

\textbf{Hybrid \acrshort{asr} experiments.} 
The hybrid \acrshort{asr} was implemented using the Kaldi toolkit~\citep{povey2011kaldi}. It consisted of a \acrfull{tdnnf} \acrshort{am}~\citep{povey2018semi}, with $12$ layers, where the hidden layers had a dimension of $1024$, and bottleneck dimension of $128$. Resnet-style bypass connections were made between each pair of consecutive \acrshort{tdnnf} layers. This \acrshort{tdnnf} \acrshort{am} was trained with the \acrfull{lfmmi} criterion~\citep{povey2016purely} for $4$ epochs. The frame-level phone alignments used as supervision for the \acrshort{am} model training were obtained by forced-alignment with a \acrshort{gmm}-\acrshort{hmm} \acrshort{am} trained beforehand. The input features were $40$-dimension high-resolution \acrshort{mfcc}s, $3$-dimension pitch features \citep{ghahremani2014pitch} and $100$-dimension i-vectors~\citep{saon2013speaker}. We used the frame length of 25ms and frame shift of 10ms.
In \acrshort{tdnnf} training, we used Kaldi's Wall Street Journal recipe\footnote{\texttt{wsj/s5/local/chain/tuning/run\_tdnn\_1g.sh}} standard hyperparameters without further tuning in our experiments. An identical \acrshort{tdnnf} structure was used for each of the experiments in this study.

The \acrshort{lm} in the hybrid \acrshort{asr} system was an n-gram phonotactic \acrshort{lm} estimated using the SRILM toolkit \citep{Stolcke02srilm--}. 
\acrshort{ipa} phonetic transcripts converted from the GlobalPhone and Babel training graphemic transcripts are used to train the phonotactic \acrshort{lm}.
We implemented uni-gram, bi-gram, and tri-gram \acrshort{lm}s to investigate the effect of imposing different amounts of phonotactic constraint on a hybrid \acrshort{asr} system in recognizing phone sequences.
To study the systems behavior under an upper bound of phonotactic constraint, we also decoded using a word-level tri-gram \acrshort{lm} and converted the word lattices to phone lattices before scoring.

The hybrid system allows us to combine \acrshort{am} and \acrshort{lm} trained on a different mix of languages. 
We leverage that possibility by combining a monolingual target language model with crosslingual \acrshort{am}, which shows the effect of adding oracle-level phonotactic knowledge to an otherwise zero-shot system \textbf{[\acrshort{rq3}]}.
This demi-oracle experiment helps us to better understand the degree of universality of \acrshort{am} representations.
We also decode using multilingual \acrshort{am} and monolingual \acrshort{lm} to understand the effect of the phonotactic model on an \acrshort{am} that has learned the target language phonetic representation.

\subsection{Phonetic inventory discovery}
\label{sec:exp:disco}

Our phonetic inventory discovery method is based on counting the phone tokens or phones, in crosslingual \acrshort{asr} transcripts. 
To convert the occurrence counts into frequencies, we divide the number of observed phone tokens by the total number of phone tokens in the transcripts. 
We checked the range of relative frequency threshold values between 1e-4 and 1e-2 and select the ones that give best performance with crosslingual \acrshort{asr} systems. We report the final threshold values in section~\ref{sec:results:discophone}.x
The optimal threshold selection might be somewhat language-dependent: in languages with a large number of phones (e.g., Cantonese) the relative phone frequencies will necessarily be lower (in expectation) than in languages with a smaller phone vocabulary.
As we don't know the inventory size when discovering inventories, we select the threshold based on the best overall performance across the 13 languages.

We test four crosslingual systems: the \acrshort{e2e} \acrshort{p} system; \acrshort{e2e} \acrshort{pt} system; and two hybrid systems with crosslingual \acrshort{lm}s -- uni-gram (weaker phonotactic constraint) and tri-gram (stronger phonotactic constraint).
To put the crosslingual \acrshort{asr} performance in perspective, we also check the performance of four other systems that have either full or partial oracle knowledge about the target language. 
The first three are the best performing systems of each kind: multilingual \acrshort{e2e} \acrshort{p}, multilingual \acrshort{e2e} \acrshort{pt}, and the monolingual hybrid system with a tri-gram word-level \acrshort{lm}.
Additionally, we test the hybrid system with a crosslingual \acrshort{am}, but using a monolingual \acrshort{lm} trained on the target language.
The last system allows us to probe the significance of knowing the target language's phonotactics in the task of phone inventory discovery -- since in actual applications we would never know them \emph{a priori}, it also reveals the limitations of this phone discovery method, as \acrshort{asr} depends on language model for accurate recognition.

We perform the phonetic inventory discovery for all thirteen languages presented in Section~\ref{sec:exp:corpora}~\textbf{[\acrshort{ma}]}.

\section{Results}
\label{sec:results}

In this section we present the experimental results.
We start with monolingual, multilingual and crosslingual (zero-shot) \acrshort{asr} in Section~\ref{sec:results:asr}. 
Then, we cast more insight on what the crosslingual system has learned by visualising its phonetic confusion matrix in Section~\ref{sec:results:phone_confusions}.
Finally, we show the results for phonetic inventory discovery in Section~\ref{sec:results:discophone}.

\subsection{ASR experiments}
\label{sec:results:asr}

\begin{table}[]
    \centering
    \begin{tabular}{lll|rr|rrr}
    \toprule
    System & AM & LM &  PER &  PTER &  INS\% &  DEL\% &  SUB\% \\
    \midrule
    \multicolumn{8}{l}{\textit{Systems with knowledge of the target language}} \\
    \midrule
    E2E P    & multi & -         &  31.8 &  29.3 &  19.2 &  29.0 &  51.8 \\
    E2E PT   & multi & -         &  -    &  32.4 &  17.8 &  31.7 &  50.5 \\
    Hybrid P & mono  & wtg-mono  &  33.6 &  44.1 &  26.6 &  28.8 &  44.6 \\
             & multi & wtg-mono  &  37.2 &  48.8 &  26.8 &  26.3 &  47.0 \\
             &       & wtg-multi &  37.3 &  48.2 &  27.7 &  24.9 &  47.5 \\
             & mono  & tg-mono   &  38.1 &  36.7 &  11.3 &  44.4 &  44.3 \\
             & multi & tg-mono   &  41.0 &  39.9 &  12.9 &  38.8 &  48.3 \\
             & mono  & bg-mono   &  43.5 &  41.0 &   8.5 &  52.1 &  39.5 \\
             & multi & tg-multi  &  43.5 &  40.9 &   9.7 &  44.9 &  45.4 \\
    E2E P    & mono  & -         &  43.9 &  39.4 &  16.1 &  28.2 &  55.7 \\
    Hybrid P & multi & bg-mono   &  45.4 &  43.4 &   9.1 &  45.9 &  45.0 \\
    E2E PT   & mono  & -         &  -    &  47.1 &  14.7 &  29.3 &  56.0 \\
    Hybrid P & multi & bg-multi  &  51.9 &  49.4 &   4.2 &  63.2 &  32.7 \\
             & mono  & ug-mono   &  54.9 &  52.6 &   3.9 &  70.3 &  25.8 \\
             & multi & ug-multi  &  63.9 &  62.6 &   1.9 &  77.5 &  20.6 \\
             &       & ug-mono   &  64.0 &  52.6 &   3.1 &  72.4 &  24.5 \\
             & cross & tg-mono   &  72.9 &  74.2 &   2.3 &  68.2 &  29.5 \\
             &       & ug-mono   &  80.9 &  82.0 &   1.1 &  79.6 &  19.3 \\
    \midrule
    \multicolumn{8}{l}{\textit{Zero-shot systems}} \\
    \midrule
    E2E PT   & cross & -         &  -    &  81.9 &   7.9 &  21.8 &  70.3 \\
    E2E P    & cross & -         &  82.1 &  83.2 &   9.7 &  20.9 &  69.4 \\
    Hybrid P & cross & bg-cross  &  83.2 &  83.6 &   1.8 &  59.4 &  38.8 \\
             &       & ug-cross  &  85.3 &  84.6 &   1.2 &  72.5 &  26.3 \\
             &       & tg-cross  &  86.1 &  93.6 &   5.2 &  40.0 &  54.7 \\
    \bottomrule
    \end{tabular}
    \caption{
        Comparison of all \acrshort{asr} systems trained in this study. 
        The results are aggregated across all 13 languages.
        For \acrshort{e2e} systems, P stands for phone symbols as recognition units, and \acrshort{pt} for phone tokens (individual \acrshort{ipa} symbols).
        For hybrid systems, ug, bg, and tg indicate (phone) uni-gram, bi-gram, or tri-gram \acrshort{lm}; and wtg stands for word tri-gram \acrshort{lm}.
        For each system, we provide the phone token error type distribution (insertion, deletion, substitution) as the percentage of the total number of phone token errors.
        We do not report \acrshort{per} for \acrshort{pt} systems, as it cannot be computed in a straight-forward manner due to ambiguities in decoding phone tokens into phones.
        }
    \label{tab:all_asr_comparison}
\end{table}

To provide the basis for answering \textbf{[\acrshort{rq1}]} and \textbf{[\acrshort{rq3}]}, we collect the results from all our \acrshort{asr} experiments in Table~\ref{tab:all_asr_comparison}.
Each row represents the aggregated scores over the 13 test languages.
For each system we show \acrshort{pter}, and for the phone-level systems also the \acrshort{per}. 
Additionally, we provide a breakdown of the percentage of insertion, deletion, and substitution errors in \acrshort{pter}.
Note that the \acrshort{e2e} \acrshort{pt} and hybrid \acrshort{p} results are based on our previous experiments in~\citep{Zelasko2020That} and in~\citep{feng2020phonotactics}, but our previous publications did not aggregate scores across all languages.  \acrshort{e2e} \acrshort{p} results have not been previously published.

Below, we present the high-level insights that we learned from performing the \acrshort{asr} experiments.

\textbf{Phones are generally preferable to phone tokens in mono- and multilingual models [\acrshort{rq1}].} 
In order to understand whether phones or phone tokens are preferable, we have to analyse their performance across mono-, multi-, and crosslingual scenarios.
We first compare the \acrshort{e2e} systems indicated with \acrshort{p} and \acrshort{pt} in Table~\ref{tab:all_asr_comparison}. 
In the group of systems that have seen the target language during training (\emph{mono}, \emph{multi}), we observe that in terms of \acrshort{pter}, the best overall system is \emph{multi} \acrshort{e2e} \acrshort{p}, closely followed by \emph{multi} \acrshort{e2e} \acrshort{pt}.
The relative improvement of \acrshort{e2e} \acrshort{p} vs \acrshort{e2e} \acrshort{pt} system is consistent when we compare the \textit{mono} variants of the \acrshort{e2e} \acrshort{p} and \acrshort{e2e} \acrshort{pt} systems.
We conclude that the \acrshort{e2e} \acrshort{p} system outperforms the \acrshort{e2e} \acrshort{pt} system. 
Possibly, the lower number of errors of the \acrshort{e2e} \acrshort{p} system is due to a larger constraint on its output token space -- the number of recognized symbol combinations is restricted due to larger vocabulary units. 
Therefore, it seems that phone-level modeling is beneficial when the test language is in-domain.

\textbf{Phone tokens might be generally preferable to phones in crosslingual models [\acrshort{rq1}].} 
Turning to the zero-shot systems section of Table~\ref{tab:all_asr_comparison}, we see that the phone-token \acrshort{e2e} scored the best at \acrshort{pter} across all crosslingual systems, with phone \acrshort{e2e} as a second and hybrid as a close third. 
We find that the \acrshort{pter} is consistently smaller in the phone-token system than in the phone system for all but two languages: Mandarin and Bengali.
Whereas the error differences are small (81.9\% vs 83.2\% and 83.6\%), this result suggests that the flexibility provided by smaller vocabulary units is helpful in crosslingual recognition. 

\textbf{Strong phonotactic model hurts crosslingual \acrshort{asr} [\acrshort{rq3}].} 
Ultimately, we want to understand the effect of phonotactics on phonetic inventory discovery. For this reason, we include and assert our previous finding from~\citep{feng2020phonotactics} here. The summary in Table~\ref{tab:all_asr_comparison} clearly shows that the \emph{tg-cross} \acrshort{lm} has the worst scores all the systems in this study. Interestingly, the \acrshort{e2e} systems scored slightly better both in terms of \acrshort{per} and \acrshort{pter}, which suggests that the \acrshort{e2e} systems might have a certain degree of robustness towards phonotactics mismatch. The best performance for \emph{cross} \acrshort{am} is observed when we introduce target language phonotactic \emph{tg-mono} \acrshort{lm} -- it approximates the lower bound of \acrshort{per}/\acrshort{pter} for a crosslingual model.

\begin{table}[th]
  \centering
  \begin{tabular}{ l r r r r r r }
    \toprule
    Language & \multicolumn{2}{c}{Mono} & \multicolumn{2}{c}{Cross} & \multicolumn{2}{c}{Multi} \\
             & PTER$^\dagger$ & PER$^\lozenge$ & PTER$^\dagger$ & PER$^\lozenge$ & PTER$^\dagger$ & PER$^\lozenge$ \\
    \midrule 
    Czech        & 26.4 & 36.4 & 65.8 & 71.5  & 8.1  & 12.5 \\
    French       & 29.5 & 30.5 & 61.7 & 62.8  & 11   & 11.5 \\
    Spanish      & 27.6 & 35.8 & 76.3 & 69.1  & 9.1  & 11.6 \\
    Mandarin     & 46.3 & 40.9 & 85.9 & 80.3  & 17.2 & 18.1 \\
    Thai         & 46   & 39.1 & 76.2 & 115.8 & 18.1 & 18.8 \\
    \midrule
    Cantonese    & 36.6 & 30.6 & 76.9 & 78.7 & 29.8 & 28.3 \\
    Bengali      & 41.2 & 34.7 & 81.4 & 74.0 & 34.6 & 34.0 \\
    Vietnamese   & 52.3 & 37.4 & 75   & 78.6 & 35.4 & 30.0 \\
    Lao          & 58.6 & 37.4 & 77.9 & 74.2 & 34.1 & 31.3 \\
    Zulu         & 41.7 & 43.3 & 77.3 & 77.5 & 35.8 & 35.4 \\
    Amharic      & 52.1 & 51.9 & 75.2 & 74.0 & 33.9 & 33.2 \\
    Javanese     & 53   & 51.4 & 99.6 & 76.1 & 41   & 38.3 \\
    Georgian     & 43.9 & 43.9 & 76.4 & 77.5 & 32.4 & 30.4 \\
    \bottomrule
  \end{tabular}
  \caption{
      \acrshort{e2e} recognition results for all 13 languages separately, for each of the 3 evaluation scenarios: Mono(lingual), Cross(lingual), Multi(lingual).
      Two systems are evaluated: phone-token system ($\dagger$) for which we report \acrshort{pter}, and phone system ($\lozenge$) for which we report \acrshort{per}. Besides the differences in label set construction, the systems are otherwise identical.
  }
  \label{tab:results:e2e_phonetokens}
\end{table}

\textbf{Phone tokens are preferable in most tone languages that we studied [\acrshort{rq1}].}
We present a per-language, side-by-side comparison of \acrshort{e2e} \acrshort{pt} systems \acrshort{pter} and \acrshort{e2e} \acrshort{p} systems \acrshort{per} in Table~\ref{tab:results:e2e_phonetokens}.
While \acrshort{per} and \acrshort{pter} cannot be directly compared, we can analyse whether they have the same trends across different languages and different training scenarios.
Most monolingual systems trained on tone languages tend to have lower \acrshort{per} than \acrshort{pter} (Mandarin, Thai, Cantonese, Vietnamese, Lao), indicating that the phone-based system is more adequate, while the reverse is true for some non-tonal language systems (Czech, French, Spanish). 
This suggests that \acrshort{ipa} recognition of tone languages benefits from the additional constraint on the possible \acrshort{ipa} symbol sequences in a low resource scenario.
Looking at the differences between the multilingual \acrshort{e2e} and monolingual \acrshort{e2e}, phone systems generally follow the same trend as phone-token systems, but the improvements in \acrshort{per} (vs.~the improvements in \acrshort{pter}) are relatively smaller. It seems that larger units (phones) impose a constraint on the network (i.e., an inductive bias) that helps monolingual systems perform better in many languages, reducing the opportunities for improvement with multilingual data.

\textbf{Both phone and phone token crosslingual systems may confuse tonal and non-tonal vowels [\acrshort{rq1}].}
         Our thirteen test languages are naturally divided into four categories, according to two binary variables: (1) Globalphone corpus versus Babel corpus, and (2) tonal versus non-tonal.  The Globalphone corpora (Czech, French, Spanish, Mandarin, and Thai) are read speech, recorded with little acoustic noise; the Babel corpora are a mixture of read and spontaneous speech, in a mixture of recording conditions.  The tonal languages include those from the Sino-Tibetan, Kra-Dai, Austronesian, and Niger-Congo language families (Mandarin, Cantonese, Thai, Lao, Vietnamese, and Zulu), while the non-tonal languages include those from the Indo-European, Afro-Asiatic, Austronesian, and Kartvelian families (Czech, French, Spanish, Bengali, Amharic, Javanese, and Georgian).  Among the Globalphone languages, Indo-European languages have considerably lower \acrshort{pter} and \acrshort{per} than tonal languages, under Monolingual, Cross-lingual, and Multilingual settings.  Thai is a particular outlier, with 115.8\% cross-lingual \acrshort{per}. {\em Post-hoc} evaluation of \acrshort{asr} transcripts suggests that the system incorrectly assumed that Thai is a non-tonal language, and exclusively recognized non-tonal phones; but note that the phone-token \acrshort{asr} did not make this mistake.  Among the Babel languages, Javanese is an outlier, with 99.6\% cross-lingual \acrshort{pter}.
       Analysis of the phonetic transcript output of the phone-token \acrshort{asr} showed that the phone-token \acrshort{asr} mistakenly treated Javanese as a tone language, appending tone modifiers to vowels, where there should be none;
       but note that the phone \acrshort{asr} did not make this mistake.
We know of no reason why the system should choose to mis-identify the language families of these two languages under these two particular test conditions.  We can note only that we have one example of a phone-based \acrshort{asr} that deleted lexical tone cross-lingually, and one example of a phone-token \acrshort{asr} that inserted lexical tone cross-lingually.



\subsubsection{General remarks}

Below, we present observations that we found interesting when analysing the \acrshort{asr} experimental results, but not directly related to our core research questions.

\textbf{Monolingual hybrid system is more robust than multilingual hybrid system.} 
We observe the best performance with \emph{mono} \acrshort{am} and word-level \emph{mono} \acrshort{lm}. 
We expected that adding multilingual phonetic transcripts to n-gram \acrshort{lm} training would not help (i.e., we expect that it acts as noise and ``steals'' the probability mass from target language n-grams during pruning). 
However, the acoustic model's degradation in \emph{multi} is not consistent with other works that have attempted multilingual training: we suspect this is due to the model sharing all the parameters (especially the output layer) between all languages; whereas \citep{Huang13,vesely2012language_independent,Kilgour2014}  observed multilingual improvements when each language had its own output layer.

\textbf{The distribution of error types shifts with training setups and the phonotactic model's strength.} 
\acrshort{e2e} \emph{multi} systems are somewhat ``balanced'' with about ~50\% substitutions, ~30\% deletions, and ~20\% insertions. 
The three best hybrid systems have a strong phonotactic constraint (i.e., they use a word-level \acrshort{lm}) and produce a similar error distribution, but with insertions and deletions equally likely.
As the phonotactic constraint weakens, we observe that the share of deletions drastically increases in the hybrid systems, achieving a 70-80\% share of all error types when unigram phone \acrshort{lm}s are used. 
The differences between the \acrshort{e2e} and hybrid models are very pronounced in the \emph{cross} scenario: in \acrshort{e2e}, there are relatively many more substitutions than insertions and deletions, while for the hybrid systems, the share of deletions is dominant at 40-72.5\%, depending on the phone \acrshort{lm}'s order. 
These findings mean that even if the error rates of the \acrshort{e2e} and hybrid models are similar, the transcripts will look rather different, with generally much shorter transcripts for the hybrid model.

\textbf{Multilingual \acrshort{e2e} is significantly more robust than monolingual \acrshort{e2e}.}
We previously found that multilingual \acrshort{e2e} provides significant gains with respect to mono in~\citep{Zelasko2020That}. Here, we observe the same trend in the \acrshort{e2e} \acrshort{p} system measured with \acrshort{per} in Table~\ref{tab:results:e2e_phonetokens}.

\subsection{Visualizing phonetic confusions}
\label{sec:results:phone_confusions}

\begin{figure*}[h]
    \begin{subfigure}{\textwidth}
      \centerline{
        \includegraphics[width=1.1\textwidth]{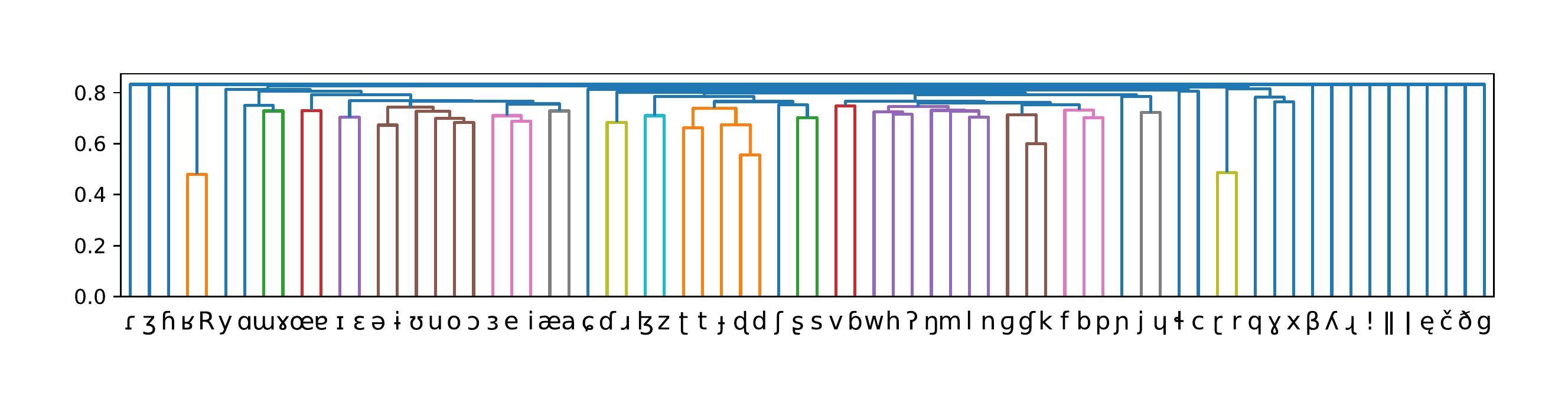}}
      \caption{Multilingual system (coloring threshold 0.75).}
      \label{fig:dendrogram:multi}
    \end{subfigure}
    \begin{subfigure}{\textwidth}
      \centerline{
        \includegraphics[width=1.1\textwidth]{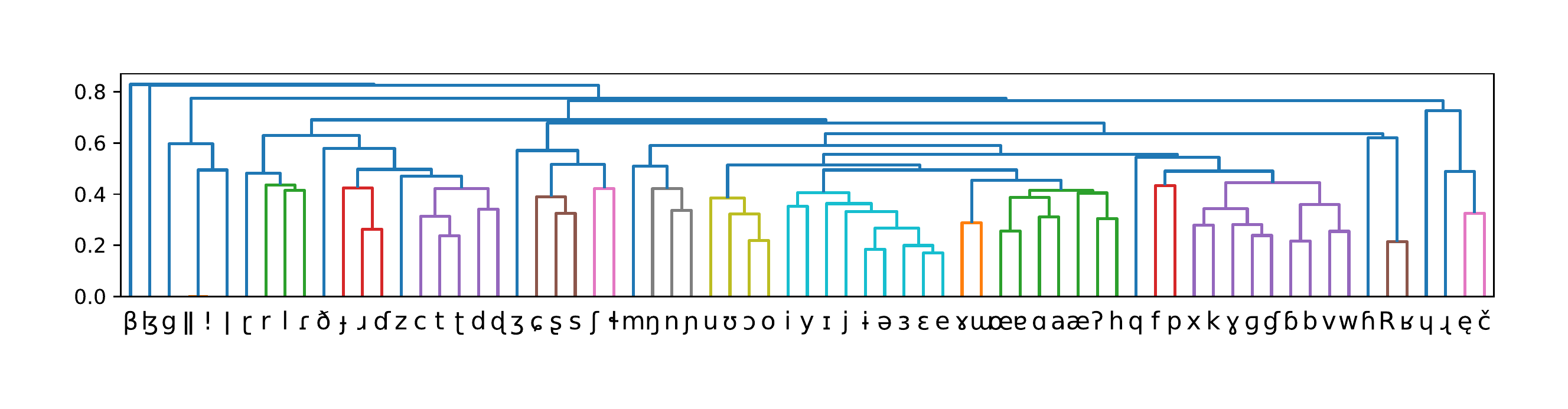}}
      \caption{Zero-shot system (coloring threshold 0.35).}
      \label{fig:dendrogram:cross}
    \end{subfigure}
  \caption{The results of the hierarchical clustering of the phone confusion matrix, presented as dendrograms. The height of each bar corresponds to a distance score (i.e., a lower bar indicates a higher similarity). We selected the threshold for cluster coloring to highlight some of the more meaningful similarities found between phones.}
  \label{fig:dendrogram}
\end{figure*}

The dendrograms resulting from the hierarchical clustering of the phone token confusions of the multilingual phone \acrshort{e2e} system are shown 
in Figure~\ref{fig:dendrogram:multi} and for the zero-shot phone \acrshort{e2e} system in Figure~\ref{fig:dendrogram:cross} \textbf{[\acrshort{rq2}]}. 
The hierarchies in the figure can be understood as groups of phone tokens that tend to have similar patterns of confusions -- the symbols that are clustered ``lower'' in the Y axis of the figure have more similarity (smaller \acrshort{jsd}).

At a  glance we see that a much higher threshold is required to cluster phone tokens in the multi system~(Figure~\ref{fig:dendrogram:multi}) than in the zero-shot system (Figure~\ref{fig:dendrogram:cross}); this is because there are fewer recognition errors in the multi system, and the clustering algorithm therefore considers its phone tokens to be more distinct.
The high error rates obtained in the zero-shot \acrshort{asr} experiments (see the \emph{Zero-shot systems} section of Table~\ref{tab:all_asr_comparison}) could suggest that the systems' outputs are close to random. However, we observe the patterns in the zero-shot system's confusions to be quite meaningful from an articulatory feature point of view  (as indicated by different phones within a cluster of the same color) and similar to the multilingual system's patterns. For instance, both the multilingual and the zero-shot systems cluster vowels separately from consonants.
The phone tokens in the zero-shot system are further clustered, with a clustering threshold of about 0.55, into approximately eight categories  defined by articulatory features, with meaningful sub-categories.
Moving from left to right through Figure~\ref{fig:dendrogram:cross}, we find that the first group contains voiced fricatives \ipa{/B,\textlyoghlig/}, a voiced plosive \ipa{/g/}, and the three clicks \ipa{/|,!,||/}.
The next group contains coronal laterals, taps, trills, plosives, and non-sibilant fricatives \ipa{/\textrtailr,r,l,\textfishhookr,D,\textbardotlessj,\textturnr,\texthtd,z,c,t,\textrtailt,d,\textrtaild/}.
The third group contains coronal sibilant fricatives \ipa{/\textyogh,\textctc,\textrtails,s,S,\textbeltl/}.
The fourth group contains the nasal consonant (\ipa{/m,\ng,n,\textltailn/}).
The fifth group contains the vowels; these are further subdivided into  non-low back rounded vowels \ipa{/u,U,\textopeno,o/}, the non-low front vowels and glides \ipa{/i,y,I,j,\textbari,\textschwa,\textrevepsilon,E,e/}, the
non-low back unrounded vowels \ipa{/\textramshorns,\textturnm/}, the low vowels \ipa{/\oe,\textturna,A,a,\ae/}, and two glottal consonants \ipa{/\textglotstop,h/}.
The sixth group contains the non-coronal obstruents, further subdivided into the pharyngeal stop \ipa{/q/}, the unvoiced labials \ipa{/f,p/}, the unvoiced velars \ipa{/x,k/}, the voiced velars \ipa{/\textgamma,g,\texthtg/}, and the voiced labials \ipa{/v,w/}.
The seventh group contains glottal and uvular fricatives and trills \ipa{/H,\textscr,\textinvscr/}.
The last group contains a  set of palatal and retroflex phones: two approximants~\ipa{/\textturnh,\textturnrrtail/}, a vowel~\ipa{/\textraising{e}/}, and an afffricate~\ipa{/\v{c}/} that are apparently sufficiently different from the other phones to be separately clustered.

  
\begin{figure*}[h]
  \centering
  \includegraphics[width=\textwidth]{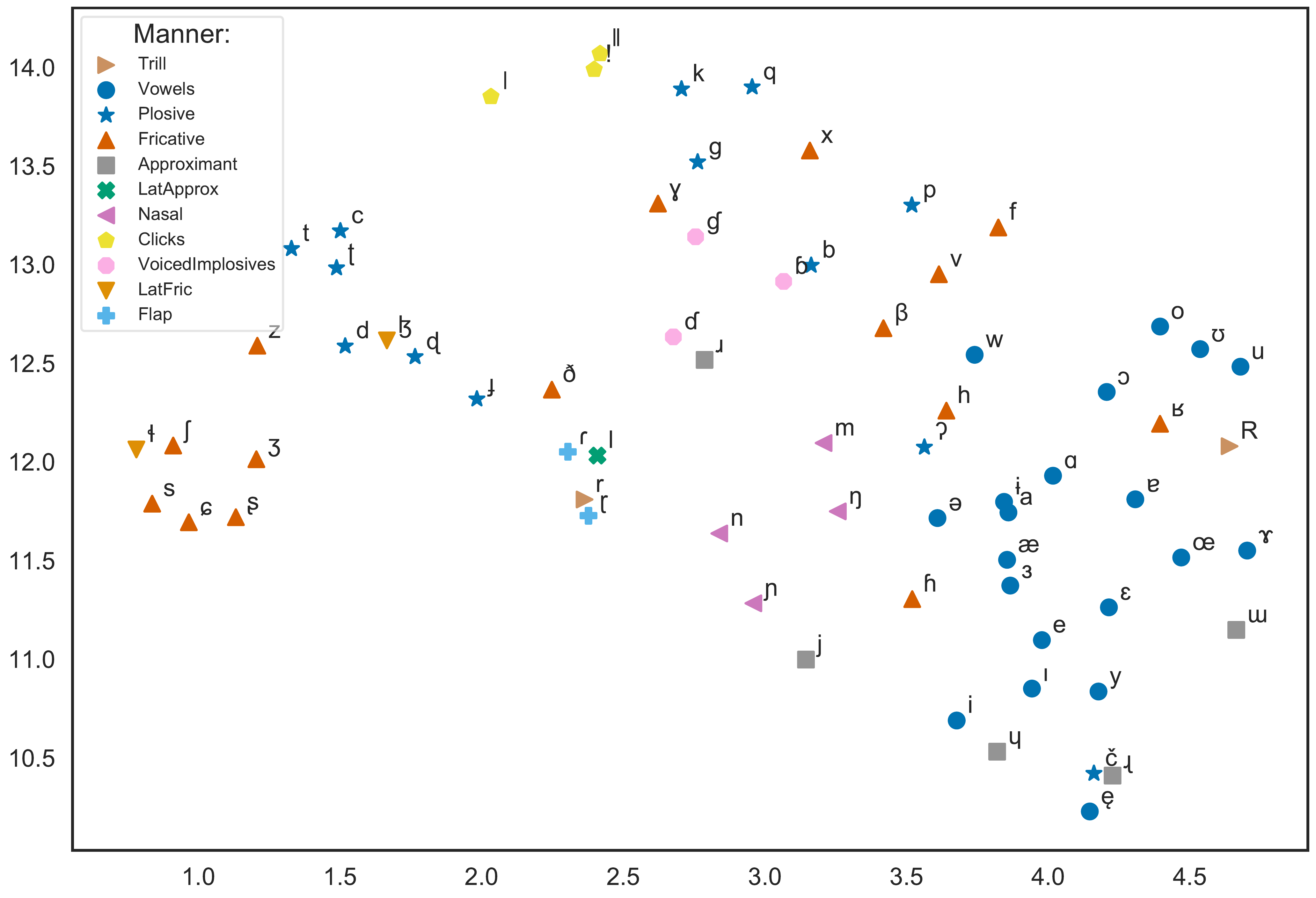}
  \caption{\acrshort{umap} projection of the phone confusion matrix for the cross-lingual \acrshort{e2e} \acrshort{p} system. Each manner of articulation class has a unique tick and a unique color to help discriminate them visually.}
  \label{fig:umap_cross}
\end{figure*}

The \acrshort{umap} projection of the crosslingual \acrshort{asr} confusion matrix is shown in
Figure~\ref{fig:umap_cross}, with each phone represented as a data point (projected from a vector describing the label categorical probability distribution) 
and colored/shaped with its corresponding manner of
articulation. The 2-D projection allows to further inspect the
learned similarities between various phones \textbf{[\acrshort{rq2}]}, and \acrshort{umap} -- unlike other projection methods such as \acrshort{tsne} -- attempts to preserve the global (in addition to local) structure in the projection.  In principle, the
phones that are frequently confused together should lie close to each
other on the projected surface.  Indeed, this display shows the same local
clustering structure that is visible in
Fig.~\ref{fig:dendrogram:cross}, but enhanced with global relationships: 
vowels are on the lower right-hand side of the plot;
moving towards left and top, there are tongue body and labial consonants,
tongue-blade consonants are on the left-hand side,
nasals are in the lower center, 
and clicks are at the top.
Among the vowels, 
back rounded vowels are at the top, 
low vowels are in the center, 
and the high vowels and back unrounded vowels are toward the bottom.  

The results in Section~\ref{sec:results:asr} show that in general the phone recognition results of the crosslingual model are rather poor. However, as these analyses of the phone confusions show, many of the phones in the target language are confused with phones that are highly similar to the target phone: phones that share many articulatory features, e.g., the vowels or nasals, are clustered together (see Figure~\ref{fig:umap_cross}). This finding resembles that of a foreign language learner, for whom one of the tasks is to learn new phonemes that are not part of their native language and consequently often not getting the articulation entirely right (yet).
It also explains the massive difference between the zero-shot and multilingual systems' performance - while the zero-shot system is able to learn some phonetic representations, the presence of the target language's data closes the gap between the learned articulation styles.

\subsection{Phonetic inventory discovery}
\label{sec:results:discophone}

\begin{table}
    \centering
    \begin{tabular}{lrrr}
        \toprule
        System & Precision & Recall & F1 \\
        \midrule
        \multicolumn{4}{l}{\textit{Zero-shot systems}} \\
        \midrule
          Hybrid, cross AM, cross phone 3-gram &       65.7 &    67.6 & 66.7 \\
          E2E, phone tokens, cross AM          &       70.1 &    65.1 & 67.5 \\
          E2E, phones, cross AM                &       69.7 &    65.6 & 67.6 \\
          Hybrid, cross AM, cross phone 1-gram &       69.7 &    67.4 & 68.6 \\
        \midrule
        \multicolumn{4}{l}{\textit{Systems with oracle knowledge}} \\
        \midrule
          Hybrid, cross AM, mono phone 3-gram  &      100.0 &    69.2 & 81.8 \\
          E2E, phone tokens, multi AM          &       99.5 &    81.9 & 89.8 \\
          E2E, phones, multi AM                &       99.7 &    82.1 & 90.1 \\
          Hybrid, mono AM, mono word 3-gram    &      100.0 &    82.8 & 90.6 \\
        \bottomrule
    \end{tabular}
    \caption{Phone token inventory discovery performance of various \acrshort{asr} systems. The phone level systems transcripts were tokenized into phone tokens for fair comparison. All systems discover phone tokens at their relative frequency of $0.004$.}
    \label{tab:inventory_comparison}
\end{table}

\begin{table}
    \centering
\begin{tabular}{llrrr}
\toprule
                                System &  Threshold &  Precision &  Recall &    F1 \\
\midrule
  \multicolumn{5}{l}{\textit{Zero-shot systems}} \\
\midrule
                E2E, phones, cross AM &     min &       14.1 &    32.9 &  19.7 \\
                                      &  0.0001 &       20.3 &    31.3 &  24.7 \\
                                      &   0.001 &       35.6 &    27.6 &  31.1 \\
                                      &   \textbf{0.002} &       43.2 &    25.8 &  \textbf{32.3} \\
                                      &   0.004 &       52.4 &    22.4 &  31.4 \\
\midrule 
 Hybrid, cross AM, cross phone 1-gram    &     min &       12.0 &    34.1 &  17.8 \\
                                         &  0.0001 &       17.5 &    32.4 &  22.7 \\
                                         &   0.001 &       36.9 &    27.9 &  31.7 \\
                                         &   \textbf{0.002} &       47.9 &    25.6 &  \textbf{33.4} \\
                                         &   0.004 &       59.9 &    22.3 &  32.5 \\
\midrule
  \multicolumn{5}{l}{\textit{Systems with oracle knowledge}} \\
\midrule
  Hybrid, cross AM, mono phone 3-gram    &     \textbf{min} &       98.0 &    43.7 &  \textbf{60.4} \\
                                         &  0.0001 &       98.1 &    42.1 &  59.0 \\
                                         &   0.001 &       99.0 &    37.0 &  53.9 \\
                                         &   0.002 &       99.0 &    34.3 &  51.0 \\
                                         &   0.004 &       98.8 &    30.3 &  46.3 \\
\midrule                   
                E2E, phones, multi AM &     min &       55.8 &    85.5 &  67.5 \\
                                      &  \textbf{0.0001} &       89.6 &    80.3 &  \textbf{84.7} \\
                                      &   0.001 &       99.9 &    62.7 &  77.1 \\
                                      &   0.002 &      100.0 &    53.2 &  69.5 \\
                                      &   0.004 &      100.0 &    42.9 &  60.1 \\
\midrule  
  Hybrid, mono AM, mono word 3-gram     &     \textbf{min} &       99.3 &    94.6 &  \textbf{96.9} \\
                                        &  0.0001 &       99.5 &    85.9 &  92.2 \\
                                        &   0.001 &      100.0 &    64.0 &  78.0 \\
                                        &   0.002 &      100.0 &    54.0 &  70.2 \\
                                        &   0.004 &      100.0 &    43.5 &  60.6 \\
   
\bottomrule
\end{tabular}
    \caption{Phone inventory discovery performance of various \acrshort{asr} systems, depending on the minimum relative frequency threshold needed for discovery. \emph{min} indicates the frequency of the least frequent phone in the transcripts of the target language.}
    \label{tab:phone_inventory_comparison}
\end{table}

The phonetic inventory discovery results, aggregated over all test languages, are presented in Table~\ref{tab:inventory_comparison} for \textbf{phone token inventory discovery} and in Table~\ref{tab:phone_inventory_comparison} for \textbf{phone inventory comparison} \textbf{[\acrshort{ma}]}. 
To score the systems, we used precision, recall, and F1-score measures: to compute them, we count the number of \acrfull{tp}: the symbols correctly assigned to the phone inventory; \acrfull{fp}: the symbols that were incorrectly indicated as being part of the target language's phone inventory; and \acrfull{fn}: the symbols that were not assigned to the predicted inventory, but should have been as they are present in the true phone inventory.
The cumulative size of phone token inventories was 448, and the cumulative size of phone inventories was 1127 (see Table~\ref{tab:inventory_per_language} for per-language details).

We start with the analysis of phone token inventory discovery in Table~\ref{tab:inventory_comparison}. We observe that both \acrshort{e2e} crosslingual \acrshort{asr} systems are on par with 67.5\% and 67.6\% F1-score, suggesting that phone and phone-token distinction is not relevant for discovering the phonetic inventory of a language \textbf{[\acrshort{rq1}]}.
Interestingly, we see that the hybrid system with a weak phonotactic model performs best (68.6\%), and the one with a strong phonotactic model performs worst (66.7\%).
This finding is consistent with our previous hypothesis that learning non-target languages' phonotactics is harmful for a zero-shot \acrshort{asr} system~\citep{feng2020phonotactics} \textbf{[\acrshort{rq3}]}. If the phonotactic model had perfect knowledge of the phonotactics of the target language, the F1-score raises to 81.8\%, which is mostly due to a perfect precision score of 100\%. The reason for the high precision is that the monolingual phonotactic model has no knowledge of those \acrshort{ipa} symbols present in the other training languages but not present in the target language. So even though the \acrshort{am} can predict these phones, they are never selected as a part of the shortest lattice path during \acrshort{asr} decoding.

If we train multilingual \acrshort{e2e} \acrshort{asr} models with knowledge of the target language, we again observe very high precision scores, and a substantial increase in recall scores as well.
So, despite lack of explicit constraint on the phonotactic information, we see an increase in F1 scores for the \acrshort{e2e} \acrshort{asr} multi models.
This is consistent with the hypothesis of \citep{feng2020phonotactics} that the encoder-decoder transformer systems are learning a representation of the target languages' phonotactics \textbf{[\acrshort{rq3}]}.
Finally, the best result of 90.6\% F1-score is obtained by a fully monolingual hybrid system with a word-level tri-gram \acrshort{lm}, likely due to its stronger phonotactic constraint imposed by the presence of a decoding graph.
Interestingly, even this system fails to recall about 17\% of the \acrshort{ipa} symbols in all languages due to \acrshort{asr} errors. 
This suggests that some phone tokens are not well modelled in the \acrshort{am} and/or are still so rare in the test material that they do not occur \textbf{[\acrshort{ma}]}.

We move onto phone inventory discovery in Table~\ref{tab:phone_inventory_comparison} \textbf{[\acrshort{ma}]}. This task is more difficult than phone token discovery, as indicated by generally lower F1 scores achieved by the zero-shot systems. This is explained by much larger sizes of the inventories, in which we expect there to be even more rare symbols than in the phone token case. The hybrid \acrshort{asr} again outperforms \acrshort{e2e} with higher F1 scores, boosted mostly by a higher precision.

Table~\ref{tab:phone_inventory_comparison} presents the results for different discovery threshold values, which reveals interesting differences between the behaviour of the \acrshort{e2e} and hybrid \acrshort{asr}. Had we fixed the best crosslingual phone discovery threshold at $0.002$ for the systems with the oracle knowledge, we'd observe a lower phone inventory discovery performance at 70.6\% for the best oracle system (i.e., the hybrid mono \acrshort{am} with mono word 3-gram). However, that threshold is too high and filters out a significant portion of the phones. Indeed, as we gradually lower the threshold towards the frequency of the rarest phone in the true inventory, we observe an F1 score that  gradually increases to  a  maximum of 96.9\% F1. 

The role of phonotactics \textbf{[\acrshort{rq3}]} in phone inventory discovery is even more pronounced than in phone token inventory discovery. The hybrid system with a crosslingual \acrshort{am} and an oracle 3-gram phone \acrshort{lm} has an F1-score almost twice that of the same hybrid system using crosslingual \acrshort{lm}. We also see that the oracle \acrshort{e2e} \acrshort{p} system scores worse than the oracle hybrid system due to false positives,  which are nearly impossible for the oracle hybrid system, because of its strong phonotactic \acrshort{lm}.

\begin{table}[]
    \centering
    \begin{tabular}{lrrrlll}
    \toprule
    {} &   TP &   FP &   FN & Precision & Recall &    F1 \\
    \midrule
    Cantonese  &   29 &    7 &    4 &      80.6 &   87.9 &  84.1 \\
    Bengali    &   26 &   14 &    7 &      65.0 &   78.8 &  71.2 \\
    Vietnamese &   29 &    9 &   13 &      76.3 &   69.0 &  72.5 \\
    Lao        &   27 &    9 &    5 &      75.0 &   84.4 &  79.4 \\
    Zulu       &   26 &   12 &   19 &      68.4 &   57.8 &  62.7 \\
    Amharic    &   24 &   14 &    7 &      63.2 &   77.4 &  69.6 \\
    Javanese   &   25 &    8 &    8 &      75.8 &   75.8 &  75.8 \\
    Georgian   &   22 &   16 &    6 &      57.9 &   78.6 &  66.7 \\
    Czech      &   24 &   11 &    6 &      68.6 &   80.0 &  73.8 \\
    French     &   19 &    7 &   23 &      73.1 &   45.2 &  55.9 \\
    Mandarin   &   15 &    7 &   21 &      68.2 &   41.7 &  51.7 \\
    Spanish    &   19 &   12 &   11 &      61.3 &   63.3 &  62.3 \\
    Thai       &   17 &    5 &   16 &      77.3 &   51.5 &  61.8 \\
    ALL        &  302 &  131 &  146 &      69.7 &   67.4 &  68.6 \\
    \bottomrule
    \end{tabular}
    \caption{Per-language breakdown of \textbf{phone token inventory discovery} performance for the zero-shot hybrid system with 1-gram \acrshort{lm} (threshold $= 0.004$).}
    \label{tab:inventory_per_language}
\end{table}

\begin{table}[]
    \centering
\begin{tabular}{lrrrlll}
\toprule
{} &   TP &   FP &   FN & Precision & Recall &    F1 \\
\midrule
Cantonese  &   19 &   44 &  116 &      30.2 &   14.1 &  19.2 \\
Bengali    &   30 &   25 &   22 &      54.5 &   57.7 &  56.1 \\
Vietnamese &   13 &   27 &  216 &      32.5 &    5.7 &   9.7 \\
Lao        &   22 &   30 &  145 &      42.3 &   13.2 &  20.1 \\
Zulu       &   29 &   23 &   40 &      55.8 &     42 &  47.9 \\
Amharic    &   29 &   14 &   31 &      67.4 &   48.3 &  56.3 \\
Javanese   &   27 &   51 &    7 &      34.6 &   79.4 &  48.2 \\
Georgian   &   26 &   28 &    9 &      48.1 &   74.3 &  58.4 \\
Czech      &   23 &   18 &   12 &      56.1 &   65.7 &  60.5 \\
French     &   21 &    7 &   24 &        75 &   46.7 &  57.5 \\
Mandarin   &   16 &   13 &   77 &      55.2 &   17.2 &  26.2 \\
Spanish    &   19 &   20 &   11 &      48.7 &   63.3 &  55.1 \\
Thai       &   15 &   14 &  128 &      51.7 &   10.5 &  17.4 \\
ALL        &  289 &  314 &  838 &      47.9 &   25.6 &  33.4 \\
\bottomrule
\end{tabular}
    \caption{Per-language breakdown of \textbf{phone inventory discovery} performance for the zero-shot hybrid system with 1-gram \acrshort{lm} (threshold $= 0.002$).}
    \label{tab:phone_inventory_per_language}
\end{table}

To investigate whether the task of automatic phonetic inventory discovery is dependent on the language and language characteristics, we investigated the phonetic inventory discovery performances per language in more detail \textbf{[\acrshort{ma}]}. Table~\ref{tab:inventory_per_language} shows the \acrfull{tp}, \acrfull{fp}, \acrfull{fn}), precision, recall, and F1 scores on the phone token inventory discovery task for the crosslingual hybrid system with 1-gram phone \acrshort{lm}, i.e., the best system from Table~\ref{tab:inventory_comparison}, for each of the 13 languages. As Table~\ref{tab:inventory_per_language} shows, there are large performance differences between languages: the difference in F1 scores between the lowest scoring language Mandarin and the highest scoring language Cantonese is 32.4\%. 

Comparing these results to Table~\ref{tab:database} shows that many of the best performing languages are, surprisingly, the languages with the highest number of unique phones (in decreasing order of number of unique phones: Vietnamese, Cantonese, Lao), with the exception of Mandarin which has the 4th highest number of unique phones; while those languages with only a few unique phones (in increasing order of number of unique phones: Georgian, Spanish, Czech) typically perform worse, but with the exception of Javanese which has only 1 unique phone but has an F1 score of 75.8\%.

Turning to Table~\ref{tab:phone_inventory_per_language}, we observe that the five languages with the worst phone inventory discovery rates are the tone languages Vietnamese, Thai, Cantonese, Lao, and Mandarin.
The low F1 scores of the tone languages are primarily caused by large false negative counts, specifically, by the inability of a cross-lingual \acrshort{asr} to recognize most tone-marked vowels.
Our phone notation treats the lexical tone as a diacritic modifier of the vowel. About two thirds of the lexical tone contours occur in only one language each, therefore about two thirds of the vowels in a tone language cannot be discovered using cross-lingual \acrshort{asr}.  
Non-tone languages are generally characterized by higher recall than precision (Bengali, Javanese, Georgian, Czech, Spanish) due to more false positives. The exceptions are French and Amharic, which tend to have the least number of false positives (along with Mandarin and Thai). In fact, French is the only language which achieves a higher F1 score in phone inventory discovery than in phone token inventory discovery.

The differences between languages are potentially due to some sounds being easier to recognize cross-linguistically than others. Phoneme inventories of languages are not created in a ``random'' fashion, rather phoneme inventories are build up in an ``ordered'' fashion \citep{Gussenhoven}. For instance, all languages have at least two of the three voiceless plosives /p, t, k/; these phonemes are considered ``simple'' sounds, i.e., sounds that are easy to produce.    The simplicity of a phoneme has many different definitions, but one definition  that is common in the literature is age of acquisition (expressed in months): It has been suggested that a
phoneme's typical age of acquisition is a measure of its articulatory
complexity~\citep{Kent92a}.  It has been demonstrated that age of
acquisition is a better predictor of articulation errors in dysarthria
than is either manner or place of articulation~\citep{Kim10d}, and a
similar ranking of the difficulty or simplicity of phones can be observed in the
transcription errors of non-native
transcribers~\citep{Jyothi2015aaai}. Romani et
al.~\citep{romani2017comparing} compared the phonemes of Italian using
three metrics: the rank-ordered age of acquisition, the frequency of
articulation errors in a population of 11 patients with apraxia of
speech, and the frequency of occurrence of the phoneme cross-linguistically.  They
found strong correlations, suggesting that, despite language-dependent
variation, there are universal tendencies for some phones to be easier
to learn and produce than others despite language-dependent
variation in the list of which phones are easier to learn and produce than others. We therefore investigated whether or not phones that are easier for humans to learn and produce are also easier for a cross-linguistic automatic speech
recognizer to correctly detect in the phone inventory of an unknown
language.

\begin{table}[]
    \centering
    \scriptsize
    \begin{tabular}{lrrrrlrrrrr}
    \toprule
    Symbol &  TP &  FP &  FN &  Count & Prec. [\%] &  Recall [\%] &  F1 [\%]
    & Phoible [\%] & AoS [\%] & AoA Rank \\
    \midrule
         m &  13 &   0 &   0 &     13 &     100.0 &       100.0 &   100.0 & 96 & 3.6 & 2 \\
         i &  13 &   0 &   0 &     13 &     100.0 &       100.0 &   100.0 & 92 & - & - \\
         j &  13 &   0 &   0 &     13 &     100.0 &       100.0 &   100.0 & 90 & 7.0 & 4 \\
         k &  13 &   0 &   0 &     13 &     100.0 &       100.0 &   100.0 & 90 & 5.6 & 1 \\
         u &  13 &   0 &   0 &     13 &     100.0 &       100.0 &   100.0 & 88 & - & - \\
         p &  13 &   0 &   0 &     13 &     100.0 &       100.0 &   100.0 & 86 & 4.5 & 2\\
         n &  13 &   0 &   0 &     13 &     100.0 &       100.0 &   100.0 & 78 & 3.7 & 2\\
         l &  13 &   0 &   0 &     13 &     100.0 &       100.0 &   100.0 & 68 & 6.6 & 1\\
         t &  13 &   0 &   0 &     13 &     100.0 &       100.0 &   100.0 & 68 & 2.2 & 1\\
         s &  13 &   0 &   0 &     13 &     100.0 &       100.0 &   100.0 & 67 & 5.8 & 2\\
         a &  12 &   1 &   0 &     12 &      92.3 &       100.0 &    96.0 & 86 & - & -\\
         o &  12 &   1 &   0 &     12 &      92.3 &       100.0 &    96.0 & 60 & - & -\\
         b &  11 &   2 &   0 &     11 &      84.6 &       100.0 &    91.7 & 63 & 15.8 & 3\\
         d &  11 &   2 &   0 &     11 &      84.6 &       100.0 &    91.7 & 46 & 16.4 & 3\\
         h &   8 &   0 &   2 &     10 &     100.0 &        80.0 &    88.9 & 56 & - & -\\
         e &  10 &   2 &   1 &     11 &      83.3 &        90.9 &    87.0 & 61 & - & -\\
         w &  10 &   2 &   1 &     11 &      83.3 &        90.9 &    87.0 & 82 & 5.9 & 4\\
         z &   7 &   0 &   3 &     10 &     100.0 &        70.0 &    82.4 & 30 & 8.4 & 6\\
         r &   8 &   5 &   0 &      8 &      61.5 &       100.0 &    76.2 & 44 & 9.3 & 3\\
         \ipa{S} &   6 &   2 &   2 &      8 &      75.0 &        75.0 &    75.0 & 37 & 13.9 & 5\\
    \bottomrule
    \end{tabular}
    \caption{Top 20 \acrshort{ipa} symbols that are successfully discovered with
      the hybrid crosslingual \acrshort{asr} system.  Phoible = \% of 3183
      phoneme inventories containing this symbol as a
      phoneme~\citep{Moran2019}.  AoS = articulation errors produced
      by patients with apraxia of speech, \% of all
      errors~\citep{romani2017comparing}. AoA = age of acquisition
      rank~\citep{romani2017comparing}.}
    \label{tab:phone_discovery_correct}
\end{table}

\begin{table}[]
    \centering
    \scriptsize
    \begin{tabular}{lrrrrlrrr}
    \toprule
    Symbol &  TP &  FP &  FN &  Count & Prec. [\%] &  Recall [\%] &  F1 [\%]
    & Phoible [\%] \\
    \midrule
         \ipa{\tone{55}} &   3 &   8 &   2 &      5 &      27.3 &        60.0 &    37.5 & 3 \\
         \ipa{A} &   0 &   8 &   3 &      3 &       0.0 &         0.0 &     0.0 & 7 \\
         \ipa{\tone{33}} &   3 &   8 &   2 &      5 &      27.3 &        60.0 &    37.5 & 10 \\
         \ipa{\tone{22}} &   2 &   8 &   2 &      4 &      20.0 &        50.0 &    28.6 & 18 \\
         \ipa{I} &   3 &   7 &   1 &      4 &      30.0 &        75.0 &    42.9 & 15 \\
         \ipa{\textbari} &   0 &   6 &   3 &      3 &       0.0 &         0.0 &     0.0 & 16 \\
         \ipa{\tone{11}} &   3 &   6 &   2 &      5 &      33.3 &        60.0 &    42.9 & 2 \\
         \ipa{\textprimstress} &   0 &   5 &   1 &      1 &       0.0 &         0.0 &     0.0 & - \\
         \ipa{U} &   2 &   5 &   0 &      2 &      28.6 &       100.0 &    44.4 & 14 \\
         \ipa{\textlengthmark} &   5 &   5 &   0 &      5 &      50.0 &       100.0 &    66.7 & - \\
         \ipa{r} &   8 &   5 &   0 &      8 &      61.5 &       100.0 &    76.2 & 44 \\
         \ipa{E} &   7 &   4 &   2 &      9 &      63.6 &        77.8 &    70.0 & 37 \\
         \ipa{\textglotstop} &   3 &   4 &   2 &      5 &      42.9 &        60.0 &    50.0 & 37 \\
         \ipa{\textschwa} &   5 &   4 &   2 &      7 &      55.6 &        71.4 &    62.5 & 22 \\
         \ipa{\ng} &   6 &   4 &   3 &      9 &      60.0 &        66.7 &    63.2 & 63 \\
         \ipa{\textsuperscript{h}} &   6 &   4 &   2 &      8 & 60.0 & 75.0 & 66.7 & - \\
         \ipa{\textfishhookr} &   0 &   3 &   1 &      1 &       0.0 & 0.0 & 0.0 & 26 \\
         \ipa{c} &   1 &   3 &   3 &      4 &      25.0 &        25.0 &    25.0 & 14 \\
         \ipa{\textopeno} &   7 &   3 &   3 &     10 &      70.0 &        70.0 &    70.0 & 35 \\
         \ipa{\textramshorns} &   0 &   2 &   2 & 2 & 0.0 & 0.0 & 0.0 & 3 \\
    \bottomrule
    \end{tabular}
    \caption{Top 20 \acrshort{ipa} symbols that are falsely discovered with the
      hybrid crosslingual \acrshort{asr} system (\textbf{false positives}).
      Phoible = \% of 3183 phoneme inventories containing this symbol
      as a phoneme~\citep{Moran2019}.}
    \label{tab:phone_discovery_falsepositive}
\end{table}

\begin{table}[]
    \centering
    \scriptsize
    \begin{tabular}{lrrrrlrrrrr}
    \toprule
    Symbol &  TP &  FP &  FN &  Count & Prec. [\%] &  Recall [\%] &  F1 [\%]
    & Phoible [\%] & AoS [\%] & AoA Rank\\    
    \midrule
         f &   6 &   0 &   7 &     13 &     100.0 &        46.2 &    63.2 & 44 & 8.5 & 4 \\
         v &   1 &   2 &   6 &      7 &      33.3 &        14.3 &    20.0 & 27 & 18.3 & 3\\
         x &   0 &   2 &   5 &      5 &       0.0 &         0.0 &     0.0 & 19 & - & -\\
         \ipa{\textltailn} &   4 &   1 &   4 & 8 & 80.0 & 50.0 & 61.5 & 42 & 23.8 & 5\\
         \ipa{\~{ }} &   0 &   0 &   3 &  3 & N/A & 0.0 & 0.0 & - & - & -\\
         \ipa{\textbari} &   0 &   6 &   3 &      3 &       0.0 &         0.0 &     0.0 & 16 & -&- \\
         \ipa{\ng} &   6 &   4 &   3 &      9 &      60.0 &        66.7 &    63.2 & 63 & -&- \\
         \ipa{\textyogh} &   4 &   1 &   3 &      7 &      80.0 &        57.1 &    66.7 & 16 & -&- \\
         \ipa{\textopeno} &   7 &   3 &   3 &     10 &      70.0 &        70.0 &    70.0 & 35 & -&- \\
         c &   1 &   3 &   3 &      4 &      25.0 &        25.0 &    25.0 & 14 & -&-\\
         \ipa{\textctc} &   0 &   1 &   3 &      3 &       0.0 &         0.0 &     0.0 & 4 & -&-\\
         \ipa{\textturnm} &   0 &   0 &   3 &      3 &       N/A &         0.0 &     0.0 & 6 & -&- \\
         y &   0 &   0 &   3 &      3 &       N/A &         0.0 &     0.0 & 6 & -&-\\
         z &   7 &   0 &   3 &     10 &     100.0 &        70.0 &    82.4 & 30 & -&- \\
         \ipa{\textrtailt} &   0 &   0 &   3 &      3 &       N/A &         0.0 & 0.0 & 16 & -&- \\
         \ipa{\textgamma} &   0 &   0 &   3 &      3 &       N/A &         0.0 &     0.0 & 14 & -&- \\
         \ipa{A} &   0 &   8 &   3 &      3 &       0.0 &         0.0 &     0.0 & 7 & -&- \\
         \ipa{H} &   0 &   1 &   2 &      2 &       0.0 &         0.0 &     0.0 & 4 & -&- \\
         \ipa{g} &   0 &   1 &   2 &      2 &       0.0 &         0.0 &     0.0 & 57 & 29.5 & 5\\
         \ipa{\textrtaild} &   0 &   0 &   2 &      2 &       N/A &         0.0 &     0.0 & 9 & -&- \\
    \bottomrule
    \end{tabular}
    \caption{Top 20 \acrshort{ipa} symbols that are missed with the hybrid
      crosslingual \acrshort{asr} system (\textbf{false negatives}). Phoible =
      \% of 3183 phoneme inventories containing this symbol as a
      phoneme~\citep{Moran2019}. AoS = articulation errors produced
      by patients with apraxia of speech, \% of all
      errors~\citep{romani2017comparing}.  AoA = age of acquisition
      rank~\citep{romani2017comparing}.}
    \label{tab:phone_discovery_falsenegative}
\end{table}

In our analyses we focused on those phone tokens (\acrshort{ipa} symbols) that are the most and
least accurately recognized symbols by the hybrid crosslingual \acrshort{asr} system with a uni-gram phone \acrshort{lm}. Table~\ref{tab:phone_discovery_correct} shows the top 20 phone tokens that are successfully discovered (i.e., the true positives). Table~\ref{tab:phone_discovery_falsepositive} shows the top 20 phone tokens that are falsely discovered (i.e., the false positives). Table~\ref{tab:phone_discovery_falsenegative} shows the top 20 phone tokens that are missed (i.e., the false negatives). In these tables, the
columns labeled ``Count'' show the number of our 13 languages that
contain this phone symbol in their phoneme inventory. The columns
labeled \acrshort{tp}, \acrshort{fp}, and \acrshort{fn} show the number of languages in which the \acrshort{ipa} symbol
was correctly detected (true positives; \acrshort{tp}), incorrectly detected (false positives; \acrshort{fp}), or incorrectly not discovered (false negatives; \acrshort{fn})
from the inventory by a cross-language speech recognizer,
respectively.  The columns labeled ``Prec,'' ``Rec,'' and ``F1'' show
the token-level precision, recall, and F1 score of each phone, using
cross-linguistic \acrshort{asr}. The three remaining columns show extrinsic
measures of the universality of each phone. The column labeled
``Phoible'' shows, as a percentage, the fraction of the 3183 phoneme
inventories at~\citep{Moran2019} that contain each phone symbol as a
phoneme. The columns labeled \acrshort{aos} and \acrshort{aoa}, both copied
from~\citep{romani2017comparing}, show the frequency of each phone's
articulation errors in the speech of people with \acrfull{aos},
and the rank-ordered \acrfull{aoa} of this phone by Italian
children, respectively; these two columns are blank for the phones
(including all phones but \ipa{/r/} in
Table~\ref{tab:phone_discovery_falsepositive}) that do not appear
in~\citep{romani2017comparing}.

Before we go deeper into the findings, first some general observations \textbf{[\acrshort{ma}]}. There are 10 symbols that have 100\% phone token inventory discovery
F1-score; 18 symbols that have $\geq80\%$ discovery F1, and 30 symbols
with $\geq60\%$ discovery F1 (see also Table~\ref{tab:phone_discovery_correct}). For reference, there are 96 unique
\acrshort{ipa} symbols, of which 56 symbols have 0\% F1.  Of those, 36
symbols are present exclusively in a single language, making their
discovery impossible with this method; 11 are present in 2-3 languages
and never hypothesized (pure false negatives; see also Table 10); 8 are present in 2-3
languages and are hypothesized, but never correctly (mix of false
negatives and false positives).  Finally, there is one symbol that has
0\% F1 and is present in more than 3 languages, i.e., \ipa{/x/} in 5
languages, where it's a false negative (see also Table~\ref{tab:phone_discovery_falsenegative}) and it is also recognized as a
false positive in two languages.

Now looking more closely at the individual phone tokens \textbf{[\acrshort{rq2}]}. Table~\ref{tab:phone_discovery_correct} shows the phone tokens with the highest \acrshort{tp} rates. Not surprisingly, these include the three ``simple'' sounds /p, t, k/, which occur in all 13 languages. Also the two nasals /m, n/ and the two approximants /l, j/, which occur in all 13 languages have a precision of 100\%. Overall, particularly, plosives, nasals, and alveolar fricatives (/s, z/) have a very high precision. This suggests that the pronunciation of these sounds, or in fact, the pronunciation of these \acrshort{ipa} symbols, is similar across the 13 languages in our dataset.

\begin{table}
  \center
  \begin{tabular}{l|rrr}
    \toprule
    & Phoible & AoS & AoA\\\hline
    Prec & 0.76** & -0.34 & -0.16 \\
    Rec & 0.67** & -0.52* & -0.52* \\
    F1 & 0.83** & -0.47 & -0.38\\
    \bottomrule
  \end{tabular}
  \caption{Pearson's correlation of cross-language phone token recognition
    scores (Precision, Recall, and F1) with extrinsic measures of
    phone token universality.  Phoible = \% of phoneme inventories
    containing the symbol~\citep{Moran2019}, AoS = articulation errors
    produced by patients with apraxia of speech, AoA = age of
    acquisition rank~\citep{romani2017comparing}. Two-sided t-test:
    *:$p<0.05$, **:$p<0.01$, with 39(Phoible) or 15(\acrshort{aos},\acrshort{aoa}) degrees
    of freedom.}
  \label{tab:extrinsic_correlations}
\end{table}

\begin{table}
    \centering
    \scriptsize
    \begin{tabular}{lllrrrr}
\toprule
        &        & Languages &  Count &  Prec. [\%] &  Recall [\%] &  F1 [\%] \\
\midrule
backness & front &    1 - 13 &     55 &       85.7 &        76.4 &    80.8 \\
        & back &    2 - 13 &     56 &       76.3 &        80.4 &    78.3 \\
        & central &    1 -  7 &     20 &       40.0 &        50.0 &    44.4 \\
        & near-back &    2 -  2 &      2 &       28.6 &       100.0 &    44.4 \\
        & near-front &    4 -  4 &      4 &       30.0 &        75.0 &    42.9 \\
height & close-mid &    1 - 12 &     26 &       81.5 &        84.6 &    83.0 \\
        & close &    3 - 13 &     51 &       86.7 &        76.5 &    81.3 \\
        & open &    3 - 12 &     15 &       57.1 &        80.0 &    66.7 \\
        & open-mid &    1 - 10 &     22 &       63.6 &        63.6 &    63.6 \\
        & mid &    1 -  7 &     15 &       55.6 &        66.7 &    60.6 \\
        & near-close &    2 -  4 &      6 &       29.4 &        83.3 &    43.5 \\
        & near-open &    1 -  1 &      2 &        0.0 &         0.0 &     0.0 \\
manner & lateral-approximant &    1 - 13 &     14 &      100.0 &        92.9 &    96.3 \\
        & approximant &    1 - 13 &     28 &       92.0 &        82.1 &    86.8 \\
        & nasal &    1 - 13 &     44 &       87.8 &        81.8 &    84.7 \\
        & plosive &    1 - 13 &     87 &       83.3 &        80.5 &    81.9 \\
        & sibilant-fricative &    2 - 13 &     43 &       88.2 &        69.8 &    77.9 \\
        & trill &    8 -  8 &      8 &       61.5 &       100.0 &    76.2 \\
        & non-sibilant-fricative &    1 - 13 &     45 &       75.0 &        33.3 &    46.2 \\
        & click &    1 -  1 &      2 &        0.0 &         0.0 &     0.0 \\
        & flap &    1 -  1 &      2 &        0.0 &         0.0 &     0.0 \\
        & implosive &    1 -  2 &      4 &        0.0 &         0.0 &     0.0 \\
        & lateral-click &    1 -  1 &      1 &        0.0 &         0.0 &     0.0 \\
        & lateral-fricative &    1 -  1 &      2 &        0.0 &         0.0 &     0.0 \\
place & bilabial &    1 - 13 &     40 &       90.2 &        92.5 &    91.4 \\
        & alveolar &    1 - 13 &     88 &       88.6 &        88.6 &    88.6 \\
        & labio-velar &   11 - 11 &     11 &       83.3 &        90.9 &    87.0 \\
        & palatal &    1 - 13 &     28 &       81.8 &        64.3 &    72.0 \\
        & palato-alveolar &    7 -  8 &     15 &       76.9 &        66.7 &    71.4 \\
        & glottal &    2 - 10 &     17 &       68.8 &        64.7 &    66.7 \\
        & velar &    1 - 13 &     40 &       72.7 &        60.0 &    65.8 \\
        & labio-dental &    1 - 13 &     21 &       77.8 &        33.3 &    46.7 \\
        & alveolo-palatal &    3 -  3 &      3 &        0.0 &         0.0 &     0.0 \\
        & dental &    1 -  1 &      2 &        0.0 &         0.0 &     0.0 \\
        & labio-palatal &    2 -  2 &      2 &        0.0 &         0.0 &     0.0 \\
        & pharyngeal &    1 -  1 &      1 &        0.0 &         0.0 &     0.0 \\
        & retroflex &    1 -  3 &      9 &        0.0 &         0.0 &     0.0 \\
        & uvular &    1 -  2 &      3 &        0.0 &         0.0 &     0.0 \\
roundness & rounded &    1 - 13 &     59 &       83.9 &        79.7 &    81.7 \\
        & unrounded &    1 - 13 &     78 &       58.5 &        70.5 &    64.0 \\
voicing & voiceless &    1 - 13 &    111 &       86.4 &        68.5 &    76.4 \\
        & voiced &    1 - 13 &    169 &       81.0 &        70.4 &    75.3 \\
\bottomrule
\end{tabular}
    \caption{A break-down of phone token inventory discovery performance by different articulatory features. The ``Languages'' column indicates the minimum and maximum number of languages that share any symbol in a given category -- e.g., in the back vowels category, every sound exists in at least 2 languages and at most 13 (all) languages. The ``Count'' column shows how many symbols there were across all languages.}
    \label{tab:artic_feats_breakdown}
\end{table}

Table~\ref{tab:phone_discovery_falsenegative} shows the phone tokens with the highest \acrshort{fn} rates.  Interestingly, fricatives are
dominant among the phone tokens with high \acrshort{fn} rates
(\ipa{/f,z,v,x,\textgamma,\textyogh/}); this point is also visible in
Table~\ref{tab:artic_feats_breakdown}, which shows that non-sibilant
fricatives and labiodentals have lower recall than the phones of any
other articulatory feature category.  Across the rows of
Tables~\ref{tab:phone_discovery_correct}
and~\ref{tab:phone_discovery_falsenegative}, the degree to which \acrshort{asr}
performance correlates with extrinsic measures of phoneme universality
is striking.  Table~\ref{tab:extrinsic_correlations} shows the
Pearson's-R correlations of cross-linguistic \acrshort{asr} Precision, Recall,
and F1 score, respectively, with the frequency of each phone token in the
languages of the world (column ``Phoible''), the frequency of phone token
error in patients with Apraxia of Speech (column ``\acrshort{aos}''), and the
phone token's Age of Acquisition among Italian children (column ``\acrshort{aoa}'').
All three \acrshort{asr} measures are correlated with cross-linguistic phone
frequency at $p<0.01$ (two-sided t-test, 39 \acrshort{dof}). Cross-linguistic
phone token recall is also correlated with articulation errors by patients
with Apraxia of Speech, and with Age of Acquisition ($p<0.05$,
two-sided t-test, 15 \acrshort{dof}).

The symbols that are most frequently falsely assigned to inventories (i.e., the highest \acrshort{fp} rates)
are shown in Table~\ref{tab:phone_discovery_falsepositive}.  Almost
half of the symbols in this table are suprasegmentals (tone symbols,
the primary stress symbol \ipa{/\textprimstress/}, and the gemination
symbol \ipa{/\textlengthmark/}).  The frequency of suprasegmentals in
this table suggests that a crosslingual \acrshort{asr} system might not be the right
tool to automatically assess the prosodic or lexical tone attributes
of a language. The remaining symbols in this table are segments that
commonly appear as allophones of phonemes written using other \acrshort{ipa}
symbols. For example, the vowels \ipa{/a/} and \ipa{/i/}, two of the
most common vowels in the languages of the world~\citep{Moran2019}, may
be implemented in many languages by their allophones
\ipa{/A,\textschwa/} and \ipa{/\textbari,I/} without loss of
intelligibility; similarly \ipa{/k/} may be implemented as \ipa{/c/}
or even \ipa{/c\textsuperscript{h}/} (as in the English words
``skit'' and ``kit,'' respectively).

Table~\ref{tab:artic_feats_breakdown} analyzes performance in the
discovery of base phones, as a function of the articulatory features
of the base phone.  Articulatory features are assigned to each symbol
according to the \acrshort{ipa} standard~\citep{international1999handbook}: vowel
symbols are characterized by backness, height, and rounding, consonant
symbols by manner, place, and voicing. 
For each category, the 
Count column is the
sum, across the inventories of all thirteen languages, of the number
of symbols of that category.  The Languages column specifies the minimum
and maximum number of languages possessing any of the symbols in that category, 
e.g., there are three distinct symbols in the [implosive] manner category,
one of which occurs in two languages, and two of which occur in just one language each,
for a total Count of 4.
Recall is the number of
true positive detections of symbols in that category, summed across
all languages, divided by the reference number of symbols in that
category, also summed across languages. Precision is the number of
true positives divided by the number of detected symbols; precision is
set to zero if there are no detections.
F1 is the harmonic mean of recall and precision.

Five manner categories, six place categories, and one vowel height
category are never detected.  All of these twelve categories are
infrequent, both in the number of distinct symbols, and in the number
of languages in which they occur: clicks, lateral clicks, and lateral
fricatives occur only in Zulu.
Other than these twelve never-detected categories, frequency of a
category is not a certain pre-requisite for high accuracy.  For
example, rounded vowels are better-detected than unrounded vowels,
despite being less frequent.  There is some tendency for categories
with low articulatory complexity to be better-detected than those with
high articulatory complexity, e.g., the best-detected manner of
articulation (lateral-approximant) is exemplified in
Table~\ref{tab:phone_discovery_correct} by the phone~\ipa{/l/}, which
has a very early age of acquisition, while the poorly-detected
non-sibilant fricative category is exemplified in
Table~\ref{tab:phone_discovery_falsenegative} by the
phones~\ipa{/f,v/} that have higher ages of acquisition.  Many of the
smaller-scale performance disparities are surprising, and perhaps
worthy of further study.  For example, close and close-mid vowels are
detected with significantly better F1 scores than are open and
open-mid vowels; we know of no theory of phonological organization
that would explain this finding.

\section{Conclusion}
\label{sec:conclusion}

To the best of our knowledge, the presented study is the first to address the problem of automatic phonetic inventory discovery using \acrshort{ipa} symbols with an application of \acrshort{asr} systems.
It concludes a series of works that began with the exploratory analysis of the universality of phonetic representations in \acrshort{e2e} \acrshort{asr}~\citep{Zelasko2020That}, and then moved onto studying the effect of phonotactic models strength on the performance of crosslingual (zero-shot) hybrid \acrshort{asr}~\citep{feng2020phonotactics}.
These works have laid the ground for addressing our ultimate goal -- i.e., phonetic inventory discovery. 

We compared the performance of two \acrshort{e2e} \acrshort{asr} systems which used phones and phone tokens as the output labels \textbf{[\acrshort{rq1}]} in Section~\ref{sec:results:asr}. We found that the phone system performs better in multi- and monolingual conditions, whereas the phone token system is preferable for crosslingual recognition. We did not find a strong effect of either label type on the F1 score performance of phone token inventory discovery. However, we found that it is significantly easier to discover phone tokens than to discover phones.

We wanted to find out empirically which phones would be easier or more difficult to automatically discover in a new language \textbf{[\acrshort{rq2}]}. To that end, we first analysed the confusions of the crosslingual phone \acrshort{e2e} \acrshort{asr} systems using Levenshtein alignments, confusion matrices, hierarchical clustering and a 2-D \acrshort{umap} projection in Section~\ref{sec:results:phone_confusions}. We presented clusters of \acrshort{ipa} symbols that are very likely to be confused with each other by a crosslingual system, and found that these clusters have a clear phonological interpretation. Furthermore, we looked closely into which phones tend to be frequently discovered, falsely hypothesized, or undiscovered in Section~\ref{sec:results:discophone}. Our findings correlate well with extrinsic indicators such as phoneme frequency in Phoible~\citep{Moran2019} or the age of acquisition~\citep{romani2017comparing}.

We previously claimed that the inability to model target language phonotactics inhibits crosslingual \acrshort{asr} performance~\citep{feng2020phonotactics}, and here we extended this claim to also involve phonetic inventory discovery with crosslingual \acrshort{asr} in Section~\ref{sec:results:discophone} \textbf{[\acrshort{rq3}]}. In particular, we found that a hybrid \acrshort{asr} system with a crosslingual \acrshort{am} (i.e., one that has never seen the target language) could exhibit much stronger phonetic inventory discovery performance if an oracle phonotactic \acrshort{lm} is available. Ideally, an acoustic model alone would be able to recognize the sounds with no phonotactic information, although that seems unlikely given the strong reliance of ASR systems (whether hybrid or \acrshort{e2e}) on the language model.

Is it possible to discover phonetic inventories with crosslingual \acrshort{asr} \textbf{[\acrshort{ma}]}? We show that to some extent, the answer is ``yes.'' Our best zero-shot system has an F1 score of 68.6\% for phone token inventory discovery, but only 33.4\% for phone inventory discovery, which presents clear continuing challenges. Most of the true positive detections are sounds that are common among many languages. However, a widespread presence is not sufficient, as common fricatives such as /f/, /v/, or /z/ tend to be harder to discover. Some sounds tend to be confused with others that differ by just a single articulatory feature, consistently with our past findings~\citep{Zelasko2020That}. We also observe that the crosslingual \acrshort{asr} struggles with determining the lexical tonality and the prosody of phones. Rare sounds, which are arguably the most interesting to discover, and languages with large inventories (especially tone languages) remain as the largest challenge for the task of automatic phonetic inventory discovery, at least for our, arguably simple, proposed method (if they can be automatically discovered at all). We believe that these challenges are adequate future work candidates.

An interesting future direction is the assessment of discovered inventory quality by training and testing an ASR system using the discovered inventory. This would help assess better how different error types (substitutions, insertions, and deletions) affect the downstream task performance. We would also like to see what patterns are observed with a larger diversity of languages -- to what extent could it reduce the number of unique phones, making their discovery possible?

We publicly share the code to reproduce our Kaldi\footnote{
\url{https://github.com/pzelasko/kaldi/tree/discophone/egs/discophone}
} and ESPnet\footnote{
\url{https://github.com/pzelasko/espnet/tree/discophone/egs/discophone/asr1}
} \acrshort{asr} experiments.



\bibliographystyle{elsarticle-num} 
\bibliography{cas-refs.bib}





\clearpage
\printglossary[type=\acronymtype]

\end{document}